\documentclass[usenatbib,epsfig]{mn2e} 

\usepackage{enumitem}
\usepackage{float}
\usepackage{epsfig}
\usepackage{epstopdf}
\usepackage{verbatim}
\usepackage{nicefrac}
\usepackage{hyperref}
\usepackage{url}
\usepackage{amsmath}
\usepackage{subfig}
\usepackage{aas_macros}
\usepackage{bm}
\usepackage[table,usenames,dvipsnames]{xcolor}
\usepackage{amsmath}
\usepackage{amssymb}
\usepackage{graphicx}
\usepackage{booktabs}

\makeatletter
\newcommand*{\rom}[1]{\expandafter\@slowromancap\romannumeral #1@}
\makeatother

\usepackage{listliketab}

\author [~Johnson et al.]
{Andrew Johnson$^{1,2}$\thanks{email: \href{mailto:asjohnson@swin.edu.au}{\nolinkurl{asjohnson@swin.edu.au}}},
Chris Blake$^1$, Jun Koda$^{1,2}$, 
Yin-Zhe Ma$^{3,4}$, \newauthor 
Matthew Colless$^{5}$,
Martin Crocce$^{6}$,
Tamara M. Davis$^{7}$, 
Heath Jones$^{8,9}$,\newauthor 
Christina Magoulas$^{9,11,15}$,
John R. Lucey$^{10}$,
Jeremy Mould$^{1}$,
Morag I. Scrimgeour$^{12,13}$,\newauthor 
Christopher M. Springob$^{2,9,14}$
 \\ \\
$^1$Centre for Astrophysics \& Supercomputing, Swinburne University of Technology, P.O. Box 218, Hawthorn, VIC 3122, Australia.\\ 
$^2$ARC Centre of Excellence for All-sky Astrophysics (CAASTRO)\\
$^3$Canadian Institute for Theoretical Astrophysics, Toronto, Canada.\\
$^{4}$Department of Physics and Astronomy, University of British Columbia, Vancouver, V6T 1Z1, BC Canada.\\
$^5$Research School of Astronomy and Astrophysics, Australian National University, Canberra, ACT 2611, Australia \\
$^{6}$Institut de Ci\`encies de l�Espai, IEEC-CSIC, Campus UAB, Facultat de Ci\`encies, Torre C5 par-2, Barcelona 08193 -Spain\\
$^{7}$School of Mathematics and Physics, University of Queensland, Brisbane, QLD 4072, Australia\\
$^{8}$School of Physics, Monash University, Clayton, VIC 3800, Australia\\
$^{9}$Australian Astronomical Observatory, P.O. Box 915, North Ryde, NSW 1670, Australia\\
$^{10}$Department of Physics, University of Durham, Durham DH1 3LE, UK\\
$^{11}$School of Physics, University of Melbourne, Parkville, VIC 3010, Australia\\
$^{12}$Department of Physics and Astronomy, University of Waterloo, Waterloo, ON, N2L 3G1, Canada\\
$^{13}$Perimeter Institute for Theoretical Physics, 31 Caroline St. N., Waterloo, ON, N2L 2Y5, Canada\\
$^{14}$International Centre for Radio Astronomy Research, M468, University of Western Australia, 35 Stirling Hwy, Crawley, WA 6009, Australia\\
$^{14}$Department of Astronomy, University of Cape Town, Private Bag X3, Rondebosch, 7701, South Africa}

\date{\today}

\pubyear{2012}

\def\beq{\begin{equation}} \def\eeq{\end{equation}}

\hypersetup{
    colorlinks=true,
    linkcolor=red,
    citecolor=blue,
} 

\title[6dFGSv: Velocity power spectrum analysis]{The 6dF Galaxy Survey: Cosmological constraints from the velocity power spectrum}

\begin{document}
\maketitle

\begin{abstract}
We present scale-dependent measurements of the normalised growth rate
of structure $f\sigma_{8}(k, z=0)$ using only the peculiar motions of galaxies. 
We use data from the 6-degree Field Galaxy Survey velocity sample (6dFGSv) 
together with a newly-compiled sample of low-redshift $(z < 0.07)$ type Ia supernovae. 
We constrain the growth rate in a series of $\Delta k \sim 0.03 h{\rm Mpc^{-1}}$ bins
to $\sim35\%$ precision, including a measurement on scales $>300 h^{-1}{\rm Mpc}$, which
represents one of the largest-scale growth rate measurement to date. We find no evidence for a
scale dependence in the growth rate, or any statistically significant variation from the growth 
rate as predicted by the {\it Planck} cosmology. Bringing all the scales together, 
we determine the normalised growth rate at $z=0$ to $\sim15\%$ in a manner
{\it independent} of galaxy bias and in excellent agreement with the constraint from
the measurements of redshift-space distortions from 6dFGS. We pay particular attention to systematic errors. 
We point out that the intrinsic scatter present in Fundamental-Plane and Tully-Fisher relations is only Gaussian 
in logarithmic distance units; wrongly assuming it is Gaussian in linear (velocity) units can bias cosmological constraints. 
We also analytically marginalise over zero-point errors in distance indicators, validate the accuracy of all our constraints using numerical simulations, 
and demonstrate how to combine different (correlated) velocity surveys using a matrix `hyper-parameter' analysis. 
Current and forthcoming peculiar velocity surveys will allow us to understand in detail the 
growth of structure in the low-redshift universe, providing strong constraints on the nature of dark energy.

\end{abstract}
\begin{keywords}
surveys, cosmology: observation, dark energy, cosmological parameters, large scale structure of the Universe
\vfill
\end{keywords}

\clearpage\section{Introduction}
\label{sec:intro}

A flat universe evolved according 
to the laws of General Relativity (GR), including a cosmological
constant $\Lambda$ and structure seeded 
by nearly scale-invariant Gaussian 
fluctuations, currently provides an excellent 
fit to a range of observations: cosmic microwave background data (CMB) \citep{Collaboration:2013qf},
baryon acoustic oscillations (BAO) \citep{2013arXiv1303.4666A, Blake:2011kl},
supernova observations \citep{2011ApJS..192....1C, Freedman:2012bs,Ganeshalingam:2013kx}, and
redshift-space distortion (RSD) measurements
\citep{2011MNRAS.415.2876B, Reid:2012ij}.
While the introduction of a cosmological constant 
term allows observational concordance by inducing a late-time 
period of accelerated expansion, its physical
origin is currently unknown. 
The inability to explain the origin of this energy density component
strongly suggests that
our current understanding of gravitation and particle 
physics, the foundations of the standard model of cosmology, 
may be significantly incomplete. 
Various mechanisms extending the standard model
have been suggested to explain this acceleration period 
such as modifying the Einstein-Hilbert action by e.g.
considering a generalised function of the Ricci scalar \citep{RevModPhys.82.451},
introducing additional matter components such as quintessence models, and
investigating the influence structure has on the large-scale evolution of the universe 
\citep{Clifton:2013kx,Wiltshire:2013vn}. 

Inhomogeneous structures in the late-time universe source gravitational 
potential wells that induce `peculiar velocities' (PVs) of galaxies, i.e., the velocity 
of a galaxy relative to the Hubble rest frame.
The quantity we measure is the line-of-sight 
PV, as this component produces Doppler distortions in the
observed redshift.
Determination of the line-of-sight 
motion of galaxies requires a redshift-independent distance estimate.
Such estimates can be performed using empirical relationships 
between galaxy properties such as the `Fundamental Plane' or
`Tully-Fisher' relation, or one can use `standard candles' such as type Ia supernovae
\citep{Colless:2001fk,Springob:2007zr, Magoulas:2009vn,Turnbull:2011qf}.
A key benefit of directly analysing PV surveys 
is that their interpretation is independent of the relation between galaxies and the 
underlying matter distribution, known as `galaxy bias' \citep[][]{gbias1}.
The standard assumptions for galaxy bias are that it is local, linear, and 
deterministic \citep{1993ApJ}; such assumptions may break down 
on small scales and introduce systematic errors in the measurement of cosmological parameters
\citep[e.g.][]{Cresswell:2008dz}. Similar issues may arise when inferring the matter 
velocity field from the galaxy velocity field: the galaxy velocity field may not move coherently 
with the matter distribution, generating a `velocity bias'. However such an effect 
is negligible given current statistical errors \citep{Desjacques:2009fk}.

Recent interest in PV surveys has been driven by the 
results of \citet{Watkins:2008kx}, which suggest that the local `bulk 
flow' (i.e. the dipole moment) of the PV field is inconsistent 
with the predictions of the standard $\Lambda$CDM model; other
studies have revealed a bulk flow more consistent with the standard model
\citep{Ma:2012vn}. PV studies were a very active field 
of cosmology in the 1990s as reviewed by 
\citet{willic} and \citet{kaiser88}. 
{Separate to the measurement of the bulk flow of local galaxies,
a number of previous studies have focused 
on extracting a measurement of the matter power spectrum in $k$-dependent bins
\citep[for example see,][]{1995ApJ...455...26J,Freudling:1999cr,2000ASPC..201..262Z,Zaroubi:2001dq,
Silberman:2001ve,Macaulay:2011bh}. This quantity is closely related to the velocity power spectrum.
Other studies have focused on directly constraining standard cosmological parameters \citep{Gordon:2007dq, Abate:2009ly}.}

The quantity we can directly measure from
the 2-point statistics of PV surveys 
is the velocity divergence power spectrum\footnote{Note 
in this analysis we will constrain the `velocity power spectrum' 
which we define as a rescaling of the more conventional velocity divergence power spectrum
(see Section \ref{sec2}).}. 
The amplitude of the velocity divergence power spectrum depends on 
the rate at which structure grows and can therefore be used to test 
modified gravity models, which have been shown to cause prominent distortions
in this measure relative to the matter power spectrum \citep{Jennings:2012zr}. 
In addition, by measuring the velocity power spectrum we are 
able to place constraints on cosmological parameters such as $\sigma_{8}$
and $\Omega_{\rm m}$
(the r.m.s of density fluctuations, at linear order, in spheres 
of comoving radius $8 h^{-1}{\rm Mpc}$; and 
the fractional matter density at $z=0$ respectively). 
Such constraints provide an interesting consistency check 
of the standard model, as the constraint on $\sigma_{8}$ measured from 
the CMB requires extrapolation from the very high redshift universe.

The growth rate of structure $f(k,a)$
describes the rate at which density perturbations 
grow by gravitational amplification. It is generically a function of the 
cosmic scale factor $a$, the comoving 
wavenumber $k$
and the growth factor $D(k,a)$; expressed as
$f(k,a) \equiv d \ln D(k,a) /d \ln a $. 
We define $\delta(k,a) \equiv \rho(k, a) /{\bar\rho(a)} -1$,
as the fractional matter over-density
and $ D(k,a) \equiv \delta(k,a)/ \delta(k,a=1)$.
The temporal dependence of the
growth rate has been readily 
measured (up to $z \sim 0.9$) by galaxy surveys
using redshift-space distortion measurements
\citep[][]{2013arXiv1312.4611B, 2011MNRAS.415.2876B, Torre:2013fu}, while 
the spatial dependence is currently only weakly constrained\footnote{
A scale dependent growth rate can be indirectly tested using the influence 
the growth rate has on the halo bias e.g. \citet{Parfrey:2010ys}.},
particularly on large spatial scales \citep[][]{2010PhRvD..81h3534B,2013JCAP...02..007D}.
{The 
observations are in fact sensitive to the `normalized
growth rate' $f(k,z)\sigma_8(z)$, which we will write as 
$f\sigma_8(k,z) \equiv f(k,z)\sigma_8(z)$.}
Recent interest in the measurement of the growth rate has been driven 
by the lack of constraining power of geometric probes on 
modified gravity models, which can generically reproduce a given 
expansion history (given extra degrees of freedom). Therefore, by combining measurements
of geometric and dynamical probes strong constraints can be
placed on modified gravity models \citep{2005PhRvD..72d3529L}.

A characteristic prediction of GR is a scale-independent growth rate,
while modified gravity models commonly
induce a scale-dependence in the growth rate.
For $f(R)$ theories of gravity this transition regime is determined 
by the Compton wavelength scale of the extra scalar degree of freedom
\citep[for recent reviews of modified gravity models see ][]{Clifton:2011nx,2010LNP...800...99T}.
Furthermore, clustering of the dark energy
can introduce a scale-dependence in the growth rate \citep{Parfrey:2010ys}.
Such properties arise in scalar field models of dark energy such as quintessence
and k-essence \citep{Caldwell:1998uq,Armendariz-Picon:2000fk}. 
The dark energy fluid is typically characterised by 
the effective sound speed
$c_{s}$ and the transition regime between clustered and smooth dark energy
is determined by the sound 
horizon \citep{Hu:2004fk}. The clustering of dark energy acts as a source 
for gravitational potential wells; therefore one finds the growth rate 
enhanced on scales above the sound horizon.
In quintessence models $c_{s}^{2} =1$; therefore
the sound horizon is equal to the particle horizon and the effect of this
transition is not measurable. Nevertheless, in models with a smaller sound speed ($c_{s}^{2} \ll 1$)
such as k-essence models, this transition may have detectable effects\footnote{
The presence of dark energy clustering requires some 
deviation from $w=-1$ in the low redshift universe.}.

Motivated by these arguments we introduce a method to measure the scale-dependence
of the growth rate of structure
using PV surveys. 
Observations from PVs are unique in this respect
as they allow constraints on the growth rate on scales inaccessible to 
RSD measurements. 
This sensitivity is a result of the relation between velocity
and density modes $v(k,z) \sim \delta(k,z)/k$ which 
one finds in Fourier 
space at linear order \citep{Dodelson-Cosmology-2003}.
The extra factor of $1/k$ gives additional weight to velocities 
for larger-scale modes relative to the density field. 
A further advantage arises because of the low redshift
of peculiar velocity surveys, namely
that the Alcock-Paczynsi effect -- transforming
the true observables (angles and redshifts) to
comoving distances -- only generates a very weak
model dependence.

A potential issue when modelling the velocity power spectrum 
is that it is known to depart from linear evolution at a larger 
scale than the density power spectrum \citep{Scoccimarro:2004ij, Jennings:2010ys}. 
We pay particular attention to  
modelling the non-linear velocity field using two loop multi-point propagators \citep{Bernardeau:2008fk}.
Additionally, we suppress non-linear contributions by smoothing the velocity field 
using a gridding procedure.
Using numerical $N$-body simulations we validate that our constraints 
contain no significant bias from non-linear effects.

For our study we use the recently compiled 6dFGSv data set along with 
low-redshift supernovae observations. 
The 6dFGSv data set represents a significant step forward in peculiar 
velocity surveys; it is the largest PV sample constructed to date
by a factor of $\sim 3$, and it covers nearly the entire southern sky.
We improve on the treatment of 
systematics and the theoretical modelling of the local velocity field, and explore 
a number of different methods to extract cosmological constraints. 
We note that the 6dFGSv data 
set will also allow constraints on the possible self-interaction 
of dark matter \citep{Linder:2013ve}, local non-Gaussianity 
\citep{Ma:2013qf}, and the Hubble flow variance
\citep{Wiltshire:2012bh}. 

The structure of this paper is as follows. In Section \ref{sec1} we 
introduce the PV surveys we analyse; Section \ref{sec2}
describes the theory behind the analysis and introduces a number 
of improvements to the modelling and 
treatment of systematics effects.
We validate our methods using numerical simulations
in Section \ref{sec:sim};
the final cosmological constraints are presented in Section \ref{sec:results}. 
We give our conclusion in Section \ref{sec:con}.

\section{Data \& Simulated Catalogues}
\label{sec1}

\subsection{6dFGS Peculiar Velocity Catalogue}

The 6dF Galaxy Survey is a combined redshift and peculiar velocity
survey that covers the whole southern sky with the exception of the region within 10 degrees of
the Galactic Plane. The survey was performed using the Six-Degree Field (6dF)
multi-fibre instrument on the UK Schmidt Telescope from
2001 to 2006. Targets were selected from the K 
band photometry of the 2MASS 
Extended Source Catalog \citep{Jarrett:2000me}.
For full details see \citet{ Jones:2004mi, Jones:2006pi, Jones:2009ff}.
To create the velocity sub-sample from the full 6dF galaxy sample the following selection
requirements were imposed:
reliable redshifts (i.e. redshift
quality Q = 3 -- 5), redshifts less than $cz < 16120$ ${\rm km~{s}^{-1}}$ in the CMB frame,
galaxies with early-type spectra, sufficiently
high signal-to-noise ratio ($S/N > 5 A^{-1}$), and 
velocity dispersions greater than the
instrumental resolution limit $\left( \sigma_{0} \ge 112 {\rm ~km~{s}^{-1}} \right)$.
This sample represents the largest and most uniformly distributed PV 
survey to date (Fig. \ref{fig:sky} top panel). The final number of galaxies with
measured PVs is 8896 and the average 
fractional distance error is $\sigma_{\rm d} = 26\%$. The redshift 
distribution for 6dFGSv is given in Fig \ref{fig:z_distribution}. 
The PVs for 6dFGSv are derived using the Fundamental Plane relation
\citep[for details of the calibration of this relation
see][]{Magoulas:2009vn, Magoulas:2012ys}. The complete 6dFGSv Fundamental 
Plane catalogue is presented in \citet{Campbell2014}.
Using the fitted Fundamental Plane relation, the final velocity catalogue is 
constructed in Springob et al. (2014). For each 
galaxy in the catalogue we determine a probability distribution 
for the quantity $\log_{10}\left( D_{\rm z} / D_{\rm H} \right)$; where $D_{\rm z}$
and $D_{\rm H}$ are respectively the `observed' comoving distance inferred from the 
observed redshift and the true comoving distance.

\subsection{Low-$z$ SNe catalogue}

\begin{figure*}
\centering
 \includegraphics[width=12cm]{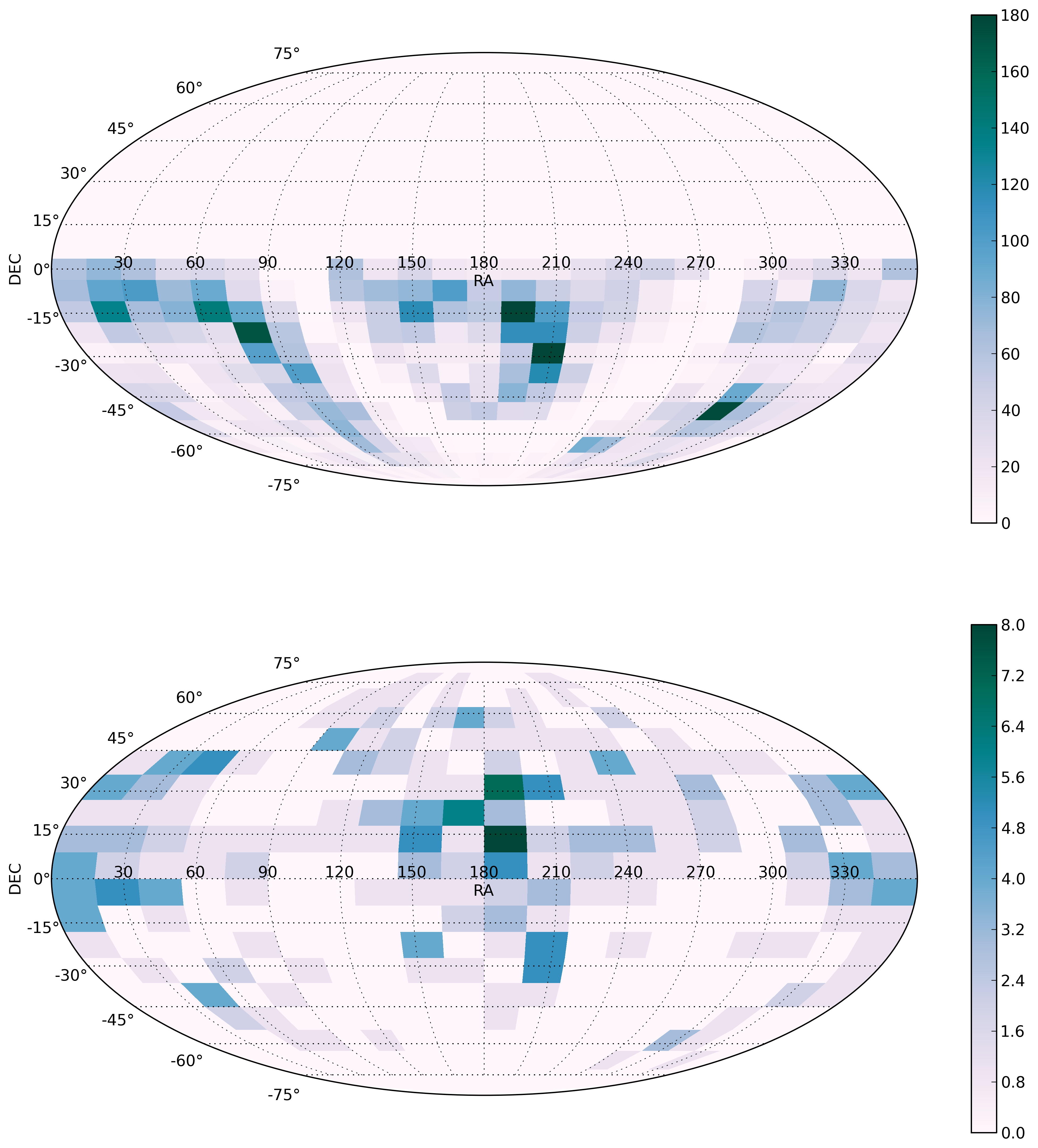}
\caption{Mollweide projection of the 6dFGSv sample (upper) and the 
low-$z$ SNe sample (lower) given in right ascension (RA) and declination (Dec) coordinates.  
We grid the RA and Dec coordinates onto a $25\times$25 grid for the upper plot and a $20\times 20$
grid for the lower plot. The colour of each cell indicates the number of galaxies with measured PVs in that 
cell; as given by the colour bars on the right.}
 \label{fig:sky}
\end{figure*}

\begin{figure*}
\centering
 \includegraphics[width=15cm]{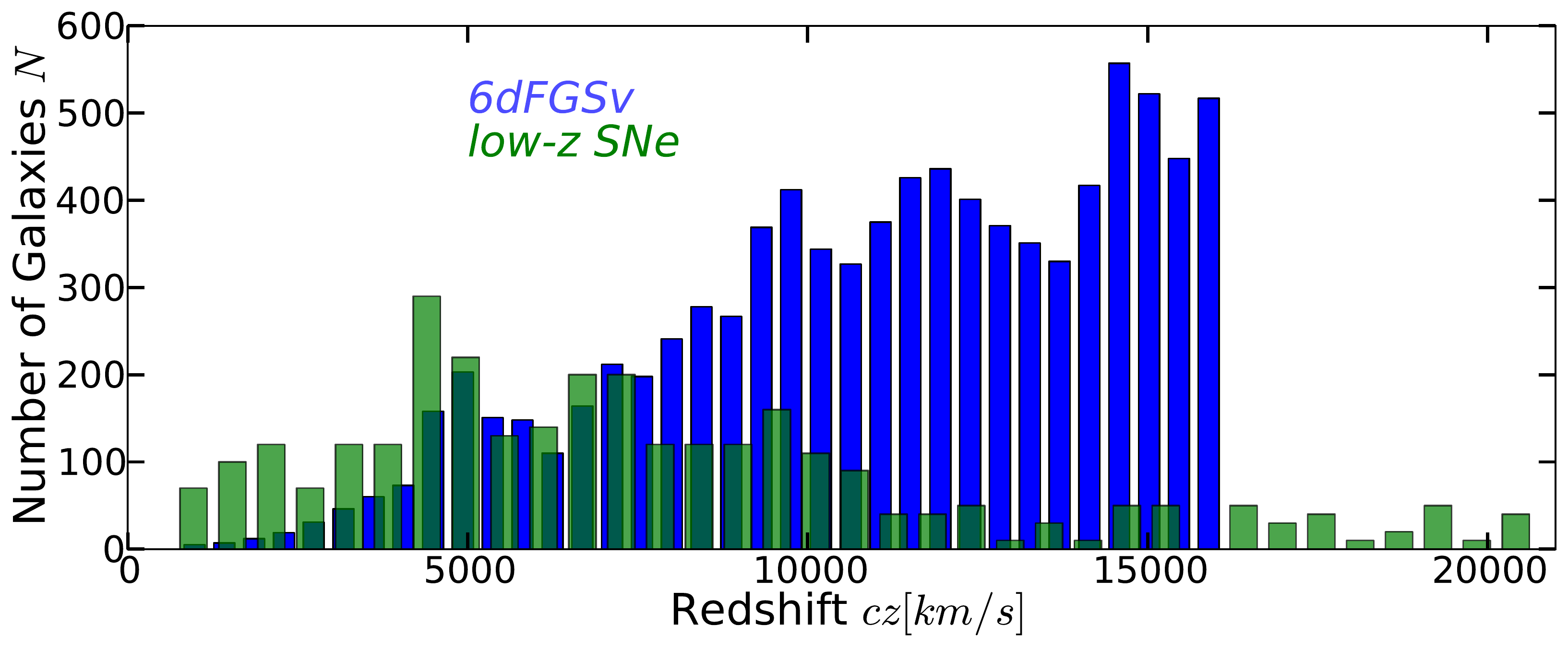}
\caption{The redshift distribution for both the 6dFGSv and low-$z$ SNe PV catalogues. Here we have scaled up the number count for the SNe sample in
each redshift bin by a factor of $10$ in order to allow the two distributions to be overplotted.}
 \label{fig:z_distribution}
\end{figure*}

To extend the velocity sample into the northern hemisphere
and cross-check the results for systematic errors, we 
construct a new homogeneous set of low-redshift Type Ia 
supernovae. 
The sample contains SNe with redshifts $z < 0.07$ and the distribution on the sky is 
given in Fig. \ref{fig:sky} (lower panel), and the redshift 
distribution is given in Fig \ref{fig:z_distribution}. 
The sample contains the following: 40 SNe from the 
Lick Observatory Supernova Search (LOSS) sample 
\citep{Ganeshalingam:2013kx}, analysed using the SALT2 light curve fitter; 
128 SNe from \citet{Tonry:2003uq}; 135 SNe from 
the `Constitution' set compiled by \citet{Hicken:2009fk}, where we choose to use the
sample reduced using the multi-color light curve shape method (MLCS) with 
their mean extinction law described by $R_{v} = 3.1$;
58 SNe in the Union sample from \citet{M.Kowalski:2008vn}\footnote{The new union2.1 data set adds no additional low-$z$ SNe.};
33 SNe from \citet{Kessler:2009ys}, where we use the sample derived using MLCS2k2 with $R_{v} = 2.18$;
and finally 26 SNe are included from the Carnegie Supernova Project (CSP) \citep{Folatelli:2009nm}.
Significant overlap exists between the samples, so for SNe with multiple 
distance modulus estimates we calculate the median 
value. This approach appears the most conservative given 
the lack of consensus between light curve reduction methods and the 
correct value of $R_{v}$; nevertheless, we find there are no significant systematic 
offsets between the different reduction methods once we correct for zero-point offsets. 
The final catalogue consists of 303 SNe with an average fractional distance error, $\sigma_{\rm d} \sim 5\%$.

We update the redshifts in these samples with the host galaxy 
redshifts in the CMB frame given in the NASA Extragalactic Database (NED), excluding SNe 
with unknown host galaxy redshifts; 
this is necessary as the quoted error in the redshift given 
for SNe data sets is similar to the typical effect that PVs have on the observed redshift.
A number of these data sets include an error 
component $\sigma_{\rm v} \sim 300$ ${\rm km~{s}^{-1}}$ accounting for peculiar motion. 
Where applicable, we removed {in quadrature} this error component of $(5/\ln(10))\sigma_{\rm v}/cz$
from the distance modulus errors. 
This component is removed
so that we can treat the samples uniformly, and in our analysis we 
treat the velocity dispersion as a free parameter.
The estimated intrinsic 
scatter in absolute magnitude $\sigma_{\rm SNe}$ is included in the error budget in all 
the samples. 
We define $\delta m \equiv \mu_{\text{obs}}(z) - \mu_{\text{Fid}}(z)$, where $\mu_{\text{Fid}}$ 
is the distance modulus calculated in a homogeneous FRW universe at redshift $z$ assuming the 
fiducial cosmology: 
$\Omega_{\rm b} =0.0489, \Omega_{\rm m}=0.3175, n_{\rm s} = 0.9624, w = -1.0, 
H_{\rm Fid} = 67 {\rm km~{s}^{-1}}\rm{Mpc}^{-1}$ \citep[motivated by][]{Collaboration:2013qf}. 

For a consistent determination of 
the line of sight PV, $S$, and the quantity $\delta m$,
the value of $H_{0}$ used to derive the prediction for the fiducial cosmology $\mu_{\rm Fid}(z)$
needs to be the same as the value assumed
during the light curve 
fitting procedure (where $\mu_{\rm obs}(z)$ is derived). 
The authors of different SNe samples have assumed 
different values of $H_{0}$ when 
deriving the distance moduli. 
Therefore before calculating $\delta m$ and the PV we correct this 
using $\Delta \mu_{i} = 5\ln(H_{0, i}/H_{\text{Fid}})$,
where $H_{0, i}$ is the assumed $H_{0}$ value in the $i^{\text{th}}$ sample and $H_{\text{Fid}}$
is the expansion rate at which we choose to normalise the sample\footnote{In the order that the SNe samples 
have been introduced the assumed velocity dispersion values are $\sigma_{\rm v} = [300,500,400,300,300,300]{\rm km~{s}^{-1}}$ and
the assumed values of the Hubble constant are $H_{0} = [70,65,65,70,65,72] {\rm km~{s}^{-1}}{\rm Mpc^{-1}}.$}.
The assumed value of $H_{\rm Fid}$ here is simply used because it is a convenient normalization.  
As $\delta m$ is a ratio of distances it is independent of the assumed value of $H_{0}$ 
(the values used to derive both distance moduli simply need to be equivalent). 

For the rest of the paper we set $H_{0} = 100h{\rm~km~s^{-1}}\rm Mpc^{-1}$. 
The line of sight PV is calculated as
\beq
S = \frac{\ln(10)}{5}\left( 1 - \frac{(1 + z)^{2}}{H(z)d_{\rm L}(z)} \right)^{-1}\delta m.
\label{eqn:1}
\eeq 
where $d_{\rm L}(z)$ is the luminosity distance and $H(z)$ the Hubble expansion rate
calculated in the fiducial model at the observed redshift $z$
(the derivation of this equation should be clear from Eq. (\ref{d_l})).

\subsection{Mock Catalogues \label{sec:mocks}}

We construct two sets of mock catalogues (\rom{1}) and (\rom{2}) using the 
GiggleZ $N$-body simulation (Poole et al. 2014). The simulation
was run inside a periodic box of $1 \text{h}^{-1}$Gpc
with $2160^{3}$ particles of mass $7.5 \times 10^{9}h^{-1}M_{\odot}$.
The simulation used the {\tt GADGET2} code \citep{Springel:2005ve}, and
haloes and sub-haloes were identified using the {\tt Subfind} algorithm \citep{Springel:2001qf}.
{ The simulation is run assuming a fiducial cosmology that is specified in Section \ref{sec:sim}.}
Using the GiggleZ simulations 10 non-overlapping realisations 
of PV surveys were constructed 
for both Mock set (\rom{1}) and (\rom{2}), with the following properties:
\begin{itemize}
\item (\rom{1}) From each central `observer' a random sample of $\sim3500$ Dark Matter 
haloes were selected within $100 h^{-1}{\rm Mpc}$ from the full sample available in the simulation
(i.e., full sky coverage). An uncertainty in the apparent magnitude of $\sigma_{\delta m} \sim 0.1$. 
was applied to each galaxy.
This corresponds to a distance error of $\sigma_{\rm d} \sim 5\%$
(viz., the approximate distance uncertainty for SNe). 
\item (\rom{2}) From each central observer $\sim8000$ Dark Matter haloes
within $150 h^{-1}{\rm Mpc}$ were selected from one hemisphere of the sky.
An error in the apparent magnitude fluctuation was introduced by 
interpolating from the observed trend for the 6dFGSv galaxies of $\sigma_{\delta m}$
with redshift. Fitting a 
simple linear relationship to the 6dFGSv data we find  
$\sigma_{\delta m} = 0.51 + 2.985z$. 
The final range of introduced observational uncertainties is 
$\sigma_{\delta m} \sim [0.5,0.75]$. 
\end{itemize}
We subsample these haloes randomly from the chosen observer volumes.
We limit the size of each hypothetical survey 
to reduce large scale correlations between the individual 
realisations, although we expect that the catalogues may still contain residual correlations
through being drawn from the same simulation. 
This situation is
more severe for Mock set (\rom{2}).
In general the purpose of 
mock set (\rom{1}) is to test the validity of our algorithms, 
various systematic effects and potential bias from non-linear effects, since
the geometry (sky coverage) of the PV survey is not important, at first order, to answer these questions. 
Mock (\rom{2}) is used as an
approximate realisation of the 6dFGSv survey.

In the mock simulations we apply a perturbation
to the PVs that is similar to the scatter induced
by observational error. The process proceeds as follows.
We place an observer in the simulation box and
extract from the simulation 
the line-of-sight velocity $S$ and true comoving distance $D_{\rm H}$
of each surrounding galaxy.
These quantities allow us to determine the observed redshift $z_{\rm obs}$, 
from $z_{\rm obs} =  (1 + z_{\rm H} )(1 + S/c) - 1$, and 
hence the observed redshift-space distance $D_{\rm z}$.
We now calculate the magnitude fluctuation $\delta m =5 \log_{10}\left( D_{\rm z} / D_{\rm H} \right)$
and apply an observational Gaussian error, using the standard deviations 
specified above. We do not attempt to include 
additional effects such as survey selection functions,
which are not required for the analysis described here.

\section{Theory \& new Methodology}
\label{sec2}

Here we discuss a number of issues, including some improvements, in the framework for 
analysing PV surveys. We pay particular attention to:

\begin{itemize}
\item{ The covariance matrix of the data (Section \ref{sec_v_cov}) } \vspace{0.10cm}
\item{ The effects of non-Gaussian observational errors and the 
	  requirement, in order to have Gaussian observational errors, 
	  to use an underlying variable that is linearly related to the
	  logarithmic distance ratio (Section \ref{sec_gaus})} \vspace{0.10cm}
\item{ The information we can
	  extract from measurements of the local velocity field using $2$-point statistics (Section \ref{sec_par})} \vspace{0.10cm}
\item{ Modelling the velocity power spectrum, including non-linear effects in redshift space (Section \ref{sec:vpower}) } \vspace{0.10cm}
\item{ Data compression using gridding methods (Section \ref{sec:gridding_met}) } \vspace{0.10cm}
\item{ Marginalization of the unknown zero-point (Section \ref{sec:zeropoint}) } \vspace{0.10cm}
\item{ Combining different correlated data sets using hyper-parameters  (Section \ref{sec:combsurveys})}
\end{itemize} 

The goal of this analysis is quantifying and modelling the degree to 
which PVs fluctuate from one part of the universe relative to other 
spatially-separated parts. The magnitude of this fluctuation
in the PV field is generated by tidal gravitational fields which are in turn
generated by the degree of departure from a homogeneous 
FRW metric and the relationship between density gradients and gravitational fields.

\begin{figure}
\includegraphics[width=8cm]{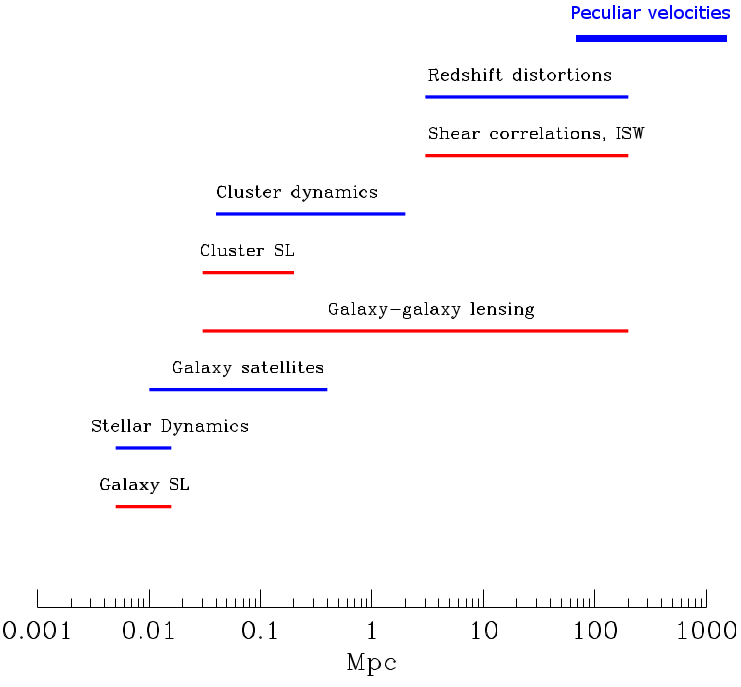}
\caption{Scales probed by different methods to constrain gravity. 
The cosmological probes shown in red lines probe gravity by its effect on 
the propagation of light i.e., weak and strong lensing (such measurements probe
the sum of the spatial and temporal gravitational potential).  
Probes that use dynamical measurements are given as blue lines (these trace the temporal 
part of the gravitational potential). PVs 
probe the largest scales of any current probe. 
Figure adapted from \protect\citet{Lombriser:2011oq}.}
\label{fig:GravityScales}
\end{figure}

We introduce a method for extracting 
scale-dependent constraints on the normalised growth rate of structure $f\sigma_{8}(z,k)$.
We emphasise
the unique ability of PV measurements to probe the growth rate 
of structure on scales that 
are not currently accessible to redshift-space 
distortion (RSD) measurements; and the
complementarity that exists between velocity surveys and RSD
measurements in constraining modified gravity theories. 
Fig. \ref{fig:GravityScales} (adapted from \citet{Lombriser:2011oq}) shows the various length scales probed
by different methods to constrain gravity. 

These methods can also be applied to larger upcoming PV surveys, such as 
the all-sky HI survey (WALLABY), the Taipan Fundamental Plane survey,
and the SDSS Fundamental Plane sample \citep{Colless:2012uq,Saulder:2013ve}
for which it will become even more crucial
to extract unbiased results with accurate error estimates.  
Furthermore the improvements considered here will be significant for other approaches 
for extracting information from velocity surveys, for example by using the 
cross-correlation between density and velocity fields.

\subsection{Velocity covariance matrix \label{sec_v_cov}}

We start with the assumption 
that the velocity field is well described by a Gaussian random field, with zero mean. 
Therefore, considering a hypothetical survey of $N$ galaxies each 
with a measured PV $S({\bf x},t) = {\bf v}({\bf x},t)\cdot {\hat r}$,
one can write down the likelihood for observing this particular field configuration as 
\beq
\mathcal {L } = \frac{1}{ |2  \pi C^{(v)} |^{1/2}}   
\exp \left ( -\frac{1}{2} \sum_{m,n} S_{m}({\bf x},t)C^{(v)  -1 }_{m n} S_{n}({ \bf x},t)\right ),
\label{eqn:likelihood}
\eeq
where {\bf v}({\bf x}, t) is the total velocity of the object evaluated at the spatial position ${\bf x}$
and time $t$, and $\hat{r}$ is a unit vector in the direction of the galaxy.
The desired (unknown) variable in this equation, which depends on the cosmological model, 
is the PV covariance matrix.
By definition
$C^{(v)}_{m n} \equiv  \left\langle S_{m} (\mathbf{x}_{m}) S_{n}
(\mathbf{x}_{n} ) \right \rangle$. 
The validity of the assumptions described above will be discussed in later sections.
{ The above approximation to the likelihood yields the probability of the velocity field 
configuration (the data $d$) given the covariance (as determined by the cosmological model $m$);}
this quantity is typically denoted ${\mathcal L} \equiv P(d | m)$. The quantity we are interested in 
extracting is the probability of the model given our 
observations of the velocity field, viz. 
$P(m | d)$. Bayes' theorem relates these two quantities as
$P(m | d) =  P(d | m) P(m) / P(d)$.
$P(d)$ can be absorbed into a normalization factor and we
assume a uniform prior (i.e, $P(m) = 1$), implying $P(m | d) \propto {\mathcal L}$.

The physical interpretation of the components of the covariance matrix are as follows:
The diagonal elements can be viewed as representing cosmic variance (later we 
add a further diagonal contribution from observational uncertainties
and non-linear contributions). As the model cosmology is changed, altering 
the degree of clustering in the low-redshift universe, the magnitude of cosmic variance changes.
The covariance between individual PVs (i.e., the off-diagonal elements) results 
from those velocities being generated by the same underlying density field.
Large wavelength Fourier density modes will have very similar phases 
for close pairs of galaxies, thus a similar gravitational force will be 
exerted on these galaxies and therefore their PVs will be 
correlated.

\begin{figure*}
\centering
 \includegraphics[width=15cm]{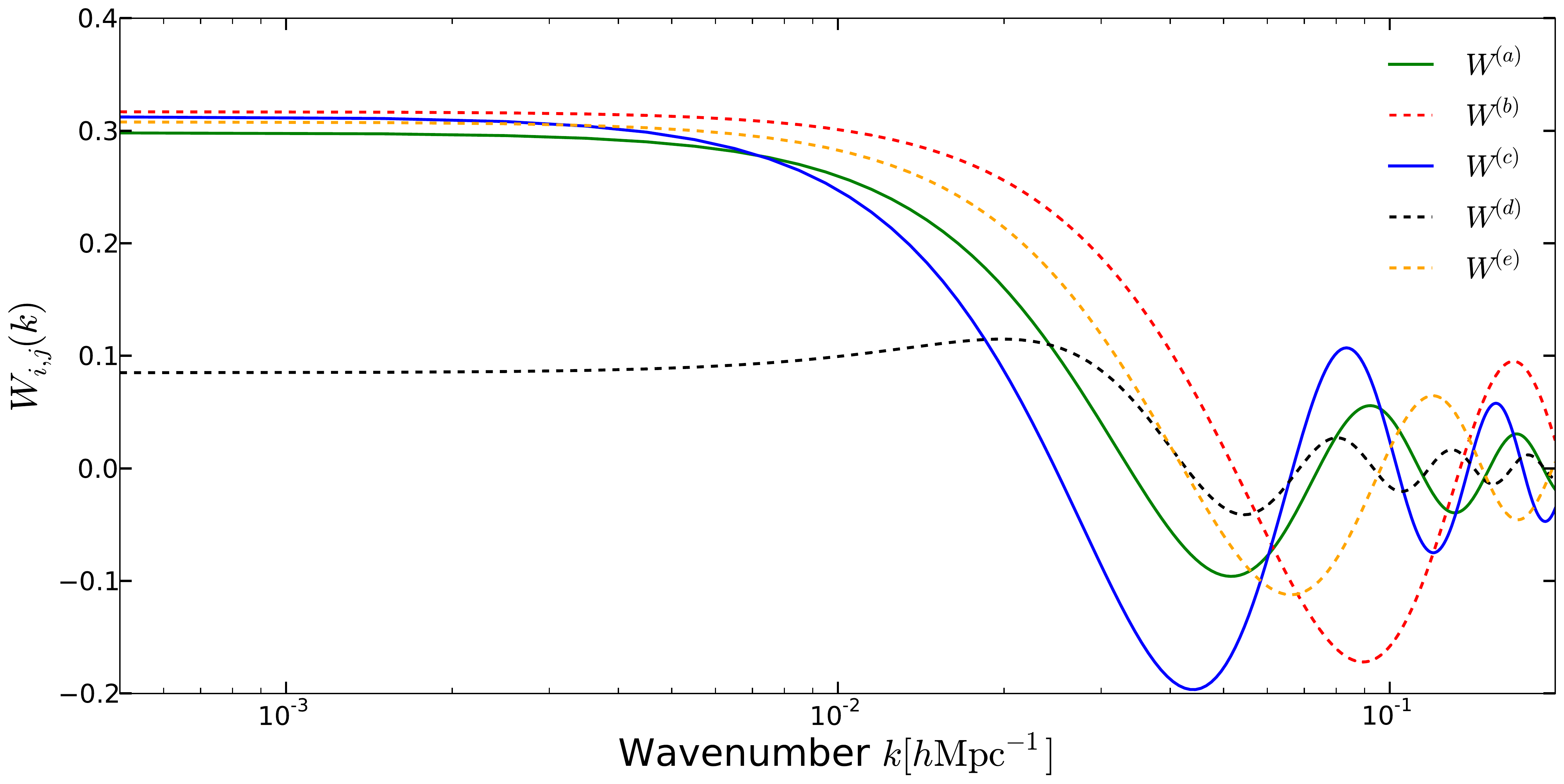}
\caption{ The window function for five pairs of galaxies in the 6dFGSv galaxy peculiar velocity catalogue. 
Large scale density fluctuations generate correlations between the PVs of pairs of galaxies, 
and the window function quantifies the wavelengths of density fluctuations that contribute to a given correlation.
Specifically, the parameters input to the above window functions are as follows: 
for $W^{(a)}$ to $W^{(e)}$ we input 
$[r_{i},r_{j},\alpha]^{(a)} = [86.6,133.7,0.393]$, $[r_{i},r_{j},\alpha]^{(b)} = [76.8,127.6,1.313]$,
$[r_{i},r_{j},\alpha]^{(c)} = [59.16,142.5,0.356]$,$[r_{i},r_{j},\alpha]^{(d)} = [51.9,91.1,0.315]$, and $[r_{i},r_{j}, \alpha]^{(e)} = [99.49,158.4,0.463]$.
The distances are all given in units of $[h^{-1}{\rm Mpc}$] and angles in radians.}
\label{fig:win_examples}
\end{figure*}

Hitherto, the covariance matrix $C^{(v)}_{m n}$ has been calculated in terms of the 
matter power spectrum, $P(k)$. We suggest that a more natural approach
is to express the covariance matrix in terms of the velocity divergence 
power spectrum. 
We define the velocity divergence
as $\theta({\bf x}, t) \equiv \nabla \cdot {\bf v}( {\bf x},t)$, therefore 
${\bf v}({\bf k}) = -i\theta({\bf k})\frac{{\bf k}}{k^{2}}$, so the velocity covariance 
matrix is given by
\begin{equation}
\begin{split}
&C^{(v)}_{m n}({\bf x}_{m}, {\bf x}_{n}) = \\
& \int \frac{d^{3}k}{(2\pi)^{3}}e^{i{\bf k}\cdot {\bf x}_{m}}  \int \frac{d^{3}k^{\prime} }{(2\pi)^{3}} e^{ -i{\bf k^{\prime}}\cdot {\bf x}_{n}} \frac{\left(\hat{x}_{m}\cdot {\bf k}\right)\left(\hat{x}_{n}\cdot{\bf k^{\prime }}\right) }{k^{\prime 2} k^{2}} \langle\theta({\bf k})\hspace{1mm}\theta^{*}({\bf k^{\prime}}) \rangle \\
&= \frac{1}{2\pi^{2}}\int d k  {\cal P}_{ \theta \theta}(k,a=1)\int \frac{d \Omega_{k}}{4\pi} e^{i{\bf k}\cdot {\left( {\bf x}_{m} - {\bf x}_{n}\right)}} \left( \hat{x}_{m}\cdot \hat{k}\right) \left( \hat{x}_{n}\cdot\hat{k}\right).
\label{vstandard}
\end{split}
\end{equation}
The simplification results from 
$\langle\theta({\bf k})\hspace{1mm}\theta^{*}({\bf k^{\prime}}) \rangle \equiv (2\pi)^{3}\delta^{3}({\bf k} - {\bf k^{\prime}}){\cal P}_{ \theta \theta}(k)$,
where ${\mathcal P}_{\theta \theta}(k)$ is the power spectrum of $\theta( {\bf x} , t )$, evaluated here at a redshift of zero.
The advantage of this derivation is that one is not required to assume the linear continuity equation.
{The angular part of the integral in Eq.~(\ref{vstandard}) defines the survey window function, explicitly} 
\beq
W(k,\alpha_{i j},r_{i},r_{j}) \equiv \int \frac{d \Omega_{k}}{4\pi} e^{i{\bf k}\cdot {\left( {\bf x}_{i} - {\bf x}_{j}\right)}} \left( \hat{x}_{i}\cdot \hat{k}\right)
\left( \hat{x}_{j}\cdot\hat{k}\right).
\label{window:def}
\eeq
The analytic form for Eq.~(\ref{window:def}) is
given in the Appendix of \citet{Ma:2010bh} as
\beq
\begin{split}
&W(k,\alpha_{i j},r_{i},r_{j})  =  1/3 \left[j_{0}(kA_{ij}) - 2j_{2}(kA_{ij})\right]\hat{r}_{i} \cdot \hat{r}_{j} \\
& + \frac{1}{A_{ij}^2}j_{2}(kA_{ij})r_{i}r_{j}\sin^{2}(\alpha_{ij})
\label{window}
\end{split}
\eeq
where $\alpha_{ij} = \cos^{-1}(\hat{r}_{i} \cdot \hat{r}_{j})$, $A_{ij} \equiv | {\bf r}_{i} - {\bf r}_{j}|$
and ${\bf r}_{i}$ is the position vector of the i$^{\text{th}}$ galaxy.
{The window function $W_{i,j}(k) \equiv W(k,\alpha_{i j},r_{i},r_{j})$ is plotted in Fig (\ref{fig:win_examples}) for a 
number of galaxy pairs in the 6dFGSv catalogue}.
For convenience we change the normalisation of the velocity divergence power spectrum and define 
the `velocity power spectrum' as
${\mathcal P}_{v v }(k) \equiv {\cal P}_{ \theta \theta}(k) /k^{2}$.
Therefore we have 
\beq
C^{(v)}_{m n} = \int \frac{d k}{ 2\pi^{2} } k^{2}{\cal P}_{ v v} (k,a=1) W(k,\alpha_{m n},r_{m},r_{n}).
\label{v_covariance}
\eeq

\subsection{The origin of non-Gaussian observational errors \label{sec_gaus}}

Observations of the Cosmic Microwave Background have shown to a very high 
degree of accuracy that the initial density fluctuations 
in the universe are Gaussian in nature, which implies that the initial velocity
fluctuations are also well-described by a Gaussian random 
field. Linear evolution of the velocity field preserves this Gaussianity, as it 
acts as a simple linear rescaling. 
This simplifying property of large scale density and velocity fields
is often taken advantage of {by approximations to the likelihood}
such as Eq.~(\ref{eqn:likelihood}), which require that the PV field, $S_{i}$,
 be accurately described by a multivariate Gaussian distribution. 
Although this is true with regards to 
cosmic variance, a crucial issue is that {\it the observational uncertainty in PV surveys is
often highly non-Gaussian in velocity units}.
In this section we describe the origin of this non-Gaussian error component, 
with particular reference to a Fundamental Plane survey;
we note our conclusions are equally valid for Tully-Fisher data sets.
Furthermore, we propose a solution to this problem and test its validity using 
numerical simulations in Section \ref{sec:sim}. 

The Fundamental Plane relation is defined as $R_{\rm e} = \sigma_{0}^{a}\langle I_{e}\rangle^{b}$,
where $R_{e}$ is the effective radius, $\sigma_{0}$ the velocity dispersion and
$\langle I_{e} \rangle$ is the mean surface brightness.  In 
terms of logarithmic quantities it is defined as $r = as + bi +c$
($r\equiv \log_{10}(R_{e})$ and $i \equiv \log_{10}(\langle I_{e}\rangle)$) where $a$ and $b$ describe the 
plane slope and $c$ defines the zero-point. 
The Fundamental Plane relation therefore is a simple linear relation when the 
relevant variables are described in logarithmic units. 
Within this parameter space (or, `Fundamental Plane space')
a $3$D elliptical Gaussian distribution provides a excellent empirical fit to the 
observed scatter of the FP variables\footnote{This scatter is generated by the 
PVs of the galaxies and the intrinsic scatter of the FP relation. Fig. 4 in
\citet{Magoulas:2012ys} shows the scatter of the FP parameters, where one can 
see the data is well described by a $3$D elliptical Gaussian (see also \citet{2003AJ....125.1817B}).}. 
Changing the distance measure $\log_{10}(R_{e})$ to a quantity not given in logarithmic units (i.e., simply $R_{e}$) 
one would find that the scatter of the new
variables can no longer be described by a simple Gaussian distribution.
This argument can be extended 
to the Tully-Fisher relation, as it has 
intrinsic scatter that appears to be modelled well by a 
Gaussian in absolute magnitude units. 

\begin{figure}
\centering
 \includegraphics[width=7.5cm]{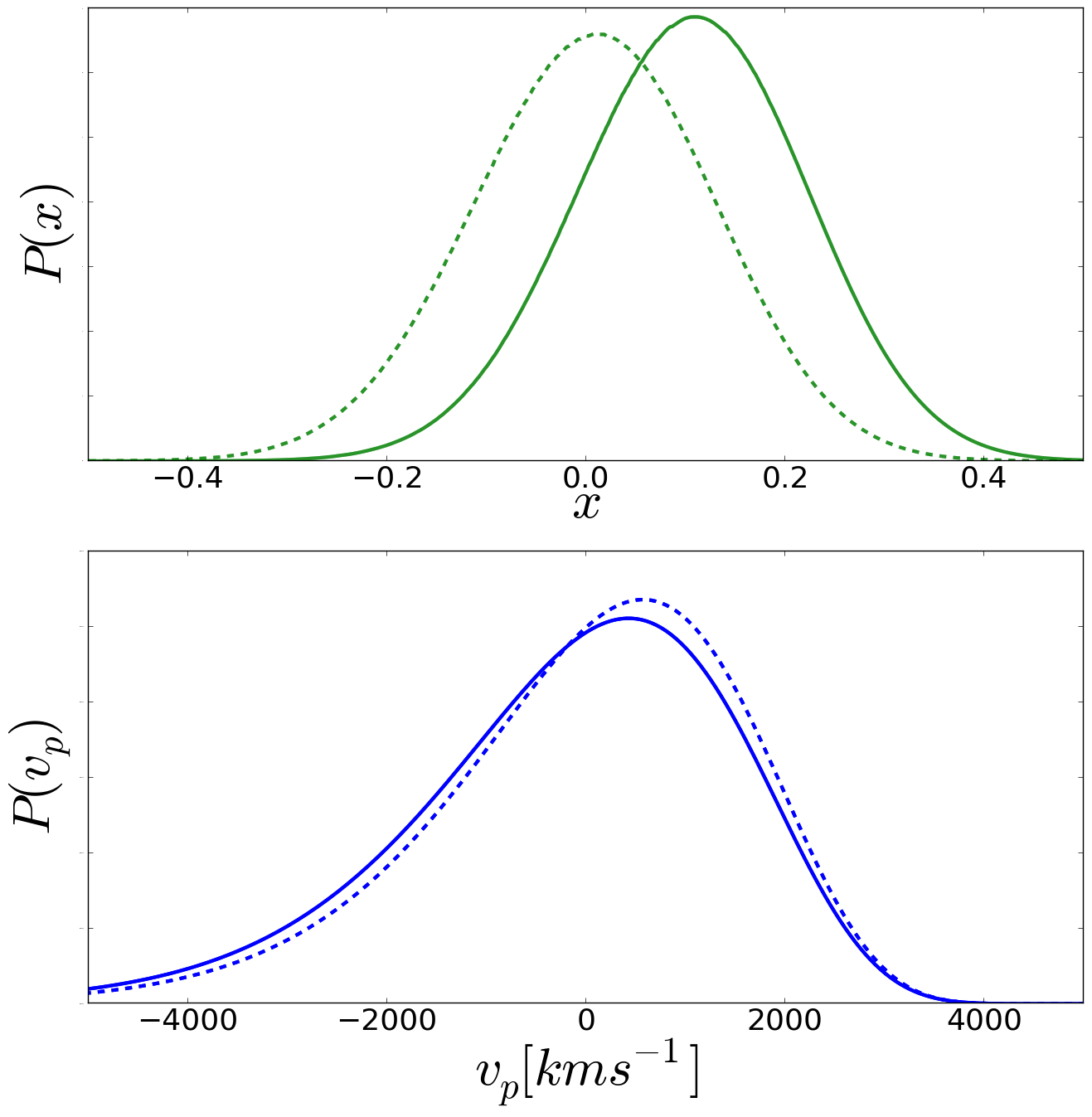}
\caption{
Probability distributions for 
$x = \log_{10} \left( D_{\rm z} / D_{\rm H} \right)$ and $v_p$ for four 6dFGS velocity sample galaxies.  We 
note that the distribution of $x$ is well-described by a Gaussian, 
whereas the distribution of $v_p$ contains significant skewness.}
\label{fig:xdis}
\end{figure}

As discussed in Springob et al. (2014) the fundamental quantity derived from the 
Fundamental Plane relation is the
probability of a given ratio between the observed effective radius (observed size) $R_{\text{z}}$ and the inferred physical 
radius (physical size) $R_{\text{H}}$ of the specific galaxy viz., $P(\log_{10}(R_{\text{z}}/R_{\text{H}}))$. 
In order to find 
the resulting probability distributions for peculiar velocities, $P(v_{p})$, in standard 
units [${\rm km~{s}^{-1}}$] from the measured quantity $P(\log_{10}(R_{\text{z}}/R_{\text{H}}))$
we need to calculate the Jacobian relating these two quantities. Firstly we can convert 
the logarithmic ratio of radii to a logarithmic ratio of comoving distances. 
Defining $x = \log_{10} \left( D_{\rm z} / D_{\rm H} \right)$,
one has
\beq
\begin{split}
P(x) &\equiv  P \left(\log_{10}{\left( D_{\rm z} / D_{\rm H} \right) }\right ) \\
& =  J(D_{\rm H},z_{\rm H} ) P(\log_{10}(R_{\rm z}/R_{\rm H})).
\label{transformationJ}
\end{split}
\eeq
{The Jacobian term needed to transform the probability distribution from a size ratio 
to a distance ratio is approximated by  (Springob et al. 2014)}
\beq
J(D_{\rm H},z_{\rm H} ) \approx  \left(1 + \frac{99.939D_{\rm H} + 0.01636D^{2}_{\rm H} } {3\times10^{5}(1 + z_{\rm H}) } \right)
\eeq
where $z_{\rm H}$ is the Hubble redshift. Any dependence on the 
assumed cosmology here will be insignificant given the low 
redshifts of the observations.
The probability distribution $P(x)$ is measured for each galaxy of the 6dFGSv
survey using Eq.~(\ref{transformationJ});
importantly this distribution is
very accurately described by a Gaussian distribution. Fig.~\ref{fig:xdis} gives
some examples for individual galaxies in the 6dFGSv sample.

We can now determine if the transformation from this distribution into the probability 
distribution for the PV (i.e.,
$P(x) \rightarrow P(v_{p})$) preserves the Gaussian nature of the distribution
or if it introduce non-Gaussianity.
The transformation between these 
two probability distributions can be accurately approximated by
\beq
P(v) = P(x)\frac{dx}{dv} \approx P(x) \frac{ (1 + z_{\rm H})^{2} } {D_{\rm H}\ln(10) c(1 + z)}\frac{d D_{\rm H} }{d z_{\rm H}}, 
\label{eqn:probcon}
\eeq
where $d D_{\rm H}/ d z_{\rm H} = c/(99.939 + 0.01636D_{\rm H})$\footnote{This result
can be derived from the approximation between comoving distance and redshift 
given in
\citet{Hogg:kx}, and is valid to $< 1\%$ within the range 
of redshift we are interested in.}. 
Applying this non-linear transformation Eq.~(\ref{eqn:probcon}) to the $P(x)$ distributions given in the $6$dFGSv sample 
we find the resulting velocity probability distributions, $P(v_{p})$, 
become significantly skewed (as shown in Fig. \ref{fig:xdis}) and hence are poorly described by a Gaussian distribution.
In Section \ref{sec:sim} we use numerical $N$-body simulations to quantify the impact
of this non-Gaussianity on cosmological parameter fits, concluding that
a measurable bias is introduced. To avoid this problem one is required to 
adopt a variable for the analysis that is linearly related to the logarithm of the 
ratio of comoving distances. 
\subsubsection{Changing variables}

The velocity variable we use is 
the apparent magnitude fluctuation, 
defined by $ \delta m(z) 
= [m(z) -\bar{m}(z)]$,
where both quantities are being evaluated at the {\it same} redshift
(the observed redshift), see \citet[][]{hui:2006fk, Davis:2010ly}.  So the fluctuation is being 
evaluated with respect to the expected apparent magnitude in {\it redshift-space}.
The over-bar here refers to the variable being evaluated 
within a homogeneous universe, i.e. in a universe with no
density gradients and as a result no PVs.
Recalling that the apparent magnitude is defined as
\beq
m = M + 5\log_{10}(d_{\rm L}(z)) + 25
\label{mag}
\eeq
where $M$ is the absolute magnitude and $d_{\rm L}(z)$ is the luminosity distance in parsecs, we find 
$\delta m(z)  = 5 x(z)$. We must now determine the covariance of magnitude fluctuations
$C^{\rm m}_{i j} \equiv \langle\delta m_{i}(z_{i}) \delta m_{j}(z_{j})\rangle$. 
The full treatment of this problem, which is effectively the derivation of the luminosity
distance in a perturbed FRW universe, includes 
a number of additional physical effects besides peculiar motion
that act to alter the luminosity distance, namely: gravitational 
lensing, the integrated Sachs-Wolfe effect, and gravitational redshift \citep{Bonvin:2006ly, Pyne:2004uq}.
For the relevant redshift range all these additional effects are currently insignificant. 
Here we focus on an intuitive derivation that captures all the relevant physics.

We first define the fractional perturbation in luminosity distance about a homogeneous universe 
as $\delta_{d_{\rm L}}(z) \equiv [d_{\rm L}(z) -\bar{d}_{\rm L}(z)] /\bar{d}_{\rm L}(z) $ and note from Eq.~(\ref{mag}) 
that $\delta m = \left(5/\ln{10} \right)\delta_{d_{\rm L}}$.
Therefore the problem is reduced to finding $C^{\rm L}_{i j} \equiv \langle \delta_{d_{\rm L}}(z_{i}) \delta_{d_{\rm L}}(z_{j})\rangle$.
The relationship between the observed flux $F$ and the intrinsic luminosity $L$ is
given by
\beq
F(z) = \frac{L}{4\pi(1 + z )^{4} }\frac{\delta \Omega_{0}}{\delta A_{\rm e}},
\eeq
where $\delta A_{\rm e}$ is the proper area of the galaxy (emitter) and $\delta \Omega_{0}$
is the observed solid angle. The angular diameter distance and the 
luminosity distance are defined as 
\beq
d_{\rm A} = \sqrt{\delta A_{e}/\delta \Omega_{0}}\hspace{4mm}	,\hspace{4mm}d_{\rm L} = d_{\rm A}(1+z)^{2},
\label{d_a_d_l}
\eeq
both of which are valid in homogeneous and 
inhomogeneous universes\footnote{ For completeness we note that the term inhomogeneous universe is 
used somewhat liberally in this section, the term should be taken to
refer to a weakly perturbed Friedmann–-Lema$\hat{\text{i}}$tre–-Robertson–-Walker geometry.
In the context of general
inhomogeneous universes the nature of the luminosity distance relation
is unknown in most cases, and other physical contributions may become 
significant.} \citep{Peebles}. 
In a homogeneous universe we have
\beq
\begin{split}
\bar{d}_{\rm A}(\bar{z}) &= \chi_{e}/(1 + \bar{z})     \\
\chi_{e}  \equiv & \chi(\bar{z}) = c\int^{\bar{z}}_{0}d z^{\prime}/ H(z^{\prime})  \\
\bar{d}_{\rm L}(\bar{z}) &= \bar{d}_{\rm A}(\bar{z})(1 + \bar{z})^{2}  
\end{split}
\eeq
where $\chi$ is the comoving distance and $H$ is Hubble's constant.
Introducing a PV component into this homogeneous system, i.e. perturbing the system,
has two effects (at first order):
\begin{itemize}
\item The redshift of the object is perturbed (via the Doppler effect). For small 
velocities (i.e., $v \ll c$), as is applicable to local motions of galaxies, the relation between the redshift 
in the homogeneous universe $\bar{z}$ and the inhomogeneous universe $z$ is given by
\beq
1+ z = (1 + \bar{z} )( 1 + \vec{v}_{\rm e} \cdot \hat{n} - \vec{v}_{0}\cdot \hat{n}),
\label{p_z}
\eeq
where $\vec{v}_{\rm e}$ is the emitting galaxy's velocity, $\vec{v}_{0}$ is the observer's velocity relative
to the CMB, and $\hat{n}$ is a unit vector in the direction of the emitter from the absorber;\\
\item The angular diameter distance is changed as a result of relativistic beaming. 
This occurs as the angle of the galaxy is
shifted by $\delta \Omega_{0} \rightarrow \delta \Omega_{0}(1 - 2\vec{v}_{0}\cdot\hat{n})$. 
The result is 
\beq
d_{\rm A}(z) = \bar{d}_{\rm A}(\bar{z})(1 + v_{0}\cdot \hat{n}).
\label{P_angular_d}
\eeq
\end{itemize} 
Using Eq.~(\ref{d_a_d_l}), Eq.~(\ref{p_z}) and Eq.~(\ref{P_angular_d}) the luminosity distance in the 
perturbed universe is given by
\beq
d_{\rm L}(z) =  \bar{d}_{\rm L}(\bar{z})(1 + 2 v_{\rm e} \cdot \hat{n}  -   v_{0} \cdot \hat{n}).
\eeq
Taylor expanding $\bar{d}_{\rm L}(z)$ about $\bar{z}$ gives \citep{hui:2006fk}
\beq
\delta_{d_{\rm L} }(z) = \frac{\delta d_{\rm L} } {d_{\rm L}} = \hat{r} \cdot \left( \vec{v}_{\rm e} - \frac{(1 + z)^{2}}{H(z)d_{\rm L}}[\vec{v}_{e} - \vec{v}_{0}]\right)
\label{d_l}
\eeq
where we work in units with $c=1$. This relation is accurate to first order in perturbation theory, ignoring other contributions.
Our Galaxy's motion is very accurately known from observations of the CMB therefore 
we can transform the observed PV to the CMB rest frame and correct for the effect of $v_{0}$\footnote{ 
We assume that the correlation between `our' motion and nearby galaxies 
is insignificant (i.e., $\langle v_{\rm e} v_{0} \rangle = 0$). 
This is justified given we are working in the CMB frame.
Any residual correlations when working in this reference frame
are introduced by the effects of relativistic beaming which is a function of our local motion.}.
Given 
$\delta m = \left(5/\ln{10} \right)\delta_{d_{\rm L}}$ and using Eq.~(\ref{v_covariance}) one finds
\begin{equation}
\begin{split}
C^{\rm m}_{i j } =  &\left( \frac{5} {\ln{10} }\right)^{2} \left( 1 - \frac{(1 + z_{i})^{2}}{H(z_{i})d_{\rm L}(z_{i})}\right)	\left( 1 - \frac{(1 + z_{j})^{2}}{H(z_{j})d_{\rm L}(z_{j})}\right) 	\\
&\int \frac{d k}{ 2\pi^{2}}k^{2} {\cal P}_{ v v}(k,a=1)W(k,\alpha_{i j},r_{i},r_{j}).
\label{cov11old}		
\end{split}
\end{equation}
In Section \ref{sec:gridding_met} 
we update the formula for the covariance matrix to account for a smoothing of the velocity field we implement;
the updated formula is given in Eq.~(\ref{cov11}).
\subsubsection{Including the intrinsic error \label{test_like}}

To complete the covariance matrix of magnitude fluctuations 
we must add the observational part of the errors, uncorrelated between 
objects. This has two different components: the error 
in the measured apparent magnitude fluctuation $\sigma_{\text{obs}}$
and a stochastic noise contribution $\sigma_{\rm v}$, which is physically 
related to non-linear contributions to the velocity \citep{Silberman:2001ve}.
The total magnitude scatter per object is given by 
\beq
\sigma^{2}_{i} = \sigma^{2}_{\rm obs} + \left( \frac{5} {\ln{10} }\right)^{2} \left( 1 -  \frac{(1 + z_{i})^{2}}{H(z_{i})d_{\rm L}(z_{i})} \right)^{2}\sigma^{2}_{\rm v},
\label{err}
\eeq
The updated posterior distribution is therefore given by 
\beq
P(\Sigma | \delta {\bf m}) = |2\pi \Sigma |^{-1/2}\exp{\left(-\frac{1}{2}{\bf \delta m}^{\text{T}} \Sigma^{-1}{\bf \delta m} \right)},
\label{like}
\eeq
where
\beq
\Sigma_{i j} \equiv C^{\rm m}_{i j } + \sigma^{2}_{i}\delta_{i j},
\label{Sig} 
\eeq
where ${\bf \delta m}$ is a vector of the observed
apparent magnitude fluctuation. {For the 
SNe sample $\sigma_{\text{obs}}$ represents both the light-curve fitting error and 
the intrinsic dispersion, as derived by the original SNe analysis.  We 
do not need to vary $\sigma_{\text{obs}}$ as a free parameter because its effect is 
degenerate with the contribution from the velocity dispersion, which we 
allow to vary.}

\subsection{Methods to extract information from the local velocity field 
\label{sec_par}}

The aim of this section is to outline the parametrisations of the velocity covariance matrix (Eq.~\ref{cov11old})
we consider, and hence the type of cosmological models we constrain.

\subsubsection{Traditional parametrisations \label{tparam}}

We first discuss two different methods already present in the literature.
 Both compare data to model by calculating a model-dependent covariance
 matrix, but they differ in the power spectrum model used to generate that 
covariance matrix.  In the first method power spectra are generated for a range
 of cosmological models (as described below), while in the second method the 
power spectra are generated in a single fiducial cosmological model, and then 
perturbed in a series of Fourier bins.  
The first method is more easily compared directly to physical models, 
while the second allows detection of generic scale-dependent effects.

Within the standard cosmological model the velocity power spectrum ${ \mathcal P}_{v v}(k)$ 
can be calculated as a function 
of the cosmological parameters $(\sigma_{8}, \Omega_{\rm m}, \Omega_{\rm b},n_{\rm s}, w, H_{0})$.
The parameters not previously described are defined as follows: 
$\Omega_{\rm b}$ is the baryon density divided by the critical density; $n_{\rm s}$
describes the slope of the primordial power spectrum; $w$ is the dark energy equation of state; and $H_0$ is 
the current expansion rate.
Current velocity data sets do not contain 
enough statistical power to constrain all these parameters, therefore we focus on the two 
most relevant parameters: $\sigma_{8}$ which describes the overall normalization and $\Omega_{\rm m}$
which controls the scale-dependence of power.
Therefore we fix $(\Omega_{\rm b} =0.0489, n_{s} = 0.9624, w = -1.0, H_{0} = 67{\rm km~{s}^{-1}}\rm{Mpc}^{-1})$ to the 
best-fitting {\it Planck} values \citep[see,][]{Collaboration:2013qf}.
Now we can parametrise the velocity power spectrum as 
$ {\cal P}_{ v v}(k) =  {\cal P}_{ v v}(k, \Omega_{\rm m},\sigma_{8})$, and
from Eq.~(\ref{cov11old}) and Eq.~(\ref{err}) we can predict the covariance matrix as a function of these 
cosmological parameters, $\Sigma = \Sigma(\Omega_{\rm m},\sigma_{8}) $, such that 
\beq
\begin{split}
&P(\Omega_{\rm m},\sigma_{8} | {\bf \delta m} ) =  \\
 &|2\pi \Sigma(\Omega_{\rm m},\sigma_{8}) |^{-1/2}\exp{\left(-\frac{1}{2}{\bf \delta m}^{\text{T}} \Sigma^{-1}(\Omega_{\rm m},\sigma_{8}){\bf \delta m} \right)}.
\label{eqn:likom}
\end{split}
\eeq
Note that the quantity $|2\pi \Sigma(\Omega_{\rm m},\sigma_{8}) |$
depends on the cosmological 
parameters, as a result we do not expect the posterior distributions 
to be exactly Gaussian. {Similar parameterisations were explored by \citet{Zaroubi:2001dq,2000ASPC..201..262Z,1995ApJ...455...26J}.}

The second method involves specifying a fiducial velocity power 
spectrum ${\cal P}_{ v v}^{\rm{Fid}}(k)$ which we choose using
the current best-fitting {\it Planck} constraints, explicitly $(\Omega_{\rm m}=0.3175, \sigma_{8}=0.8344,
\Omega_{\rm b}=0.0489, n_{\rm s}=0.9624, w=-1.0, \text{H}_{0}=67  \rm{ km~{s}^{-1}}\rm{Mpc}^{-1} )$.
The power spectrum is now separated into bins in Fourier space and
a free parameter $A_{i}$ is introduced and allowed to scale the 
`power' within the given $k$ range of a bin. 
One can hence constrain the amplitude of the velocity power spectrum 
in $k$-dependent bins. 
This parameterisation is similar in nature to that explored in  
\citet{Macaulay:2011bh} and \citet{Silberman:2001ve}, although the specifics of the
implementation are somewhat different. 
This approach is more model-independent than the first parametrisation
because it allows more freedom in the shape of the velocity power
spectrum.  
Considering a case with $N$ different bins, we define the centre of the 
 i$^{\text{th}}$ bin as $k^{\text{cen}}_{i}$ and the bin width as
$\Delta_{i} \equiv \left(k^{\text{max}}_{i} - k^{\text{min}}_{i}\right)$.
We define 
\beq
\begin{split}
&\Pi(k, \Delta_{i},  k^{\text{cen}}_{i}) \equiv \\
&{\mathcal H}(k -(k^{\text{cen}}_{i} - \Delta_{i}/2)) - {\mathcal H}(k - \left(k^{\text{cen}}_{i} + \Delta_{i}/2\right)),
\end{split}
\eeq
where ${\mathcal H}(x)$ is a Heaviside step function, so
$\Pi(k, k^{\text{cen}}_{i}, \Delta_{i})$ is equal to one if $k$ is in the $i^{\text{th}}$ bin and zero otherwise.
Including the
free parameters $A_{i}$ which scale the amplitude of the velocity power spectrum within each bin, 
the {\it scaled} velocity power spectrum is given by\footnote{Note we have by definition
\beq
\begin{split}
{\cal P}_{ v v}^{\text{Fid}}(k) &= {\cal P}_{ v v}^{\text{Fid}}(k)\Pi(k,\Delta_{1},k^{\text{cen}}_{1}) +  {\cal P}_{v v}^{\text{Fid}}(k)\Pi(k,\Delta_{2}, k^{\text{cen}}_{2})\hspace{1mm}\\
& ... + \hspace{1mm}{\cal P}_{ v v }^{\text{Fid}}(k)\Pi(k,\Delta_{N}, k^{\text{cen}}_{N}).
\end{split}
\nonumber
\eeq}
\beq
\begin{split}
 {\cal P}_{ v v}^{\text{Scaled}}(k) \equiv & A_{1}{\cal P}_{ v v}^{\text{Fid}}(k)\Pi(k,\Delta_{1}, k^{\text{cen}}_{1}) \hspace{1mm} \\
&+  A_{2}{\cal P}_{ v v}^{\text{Fid}}(k)\Pi(k, \Delta_{2}, k^{\text{cen}}_{2})\\
 &... + A_{N}{\cal P}_{ v v}^{\text{Fid}}(k)\Pi(k,\Delta_{N}, k^{\text{cen}}_{N}).
\end{split}
\label{eqn:defA}
\eeq
The free parameters $A_{i}$ do not have any $k$-dependence, and as a result
one finds
\beq
\begin{split}
\int \frac{d k}{2\pi^{2}}k^{2} {\cal P}_{ v v }^{\text{Scaled}}(k) W(k,\alpha_{12},r_{1},r_{2}) 	&= \\
\sum_{i=1}^{N} A_{i} \int^{k^{\text{cen}}_{i} + \Delta_{i}/2}_{k^{ \text{cen} }_{i} -\Delta_{i}/2} \frac{d k}{2\pi^{2}}& k^{2}{\cal P}_{ v v}^{\text{Fid}}(k) W(k,\alpha_{12},r_{1},r_{2})
\label{cov12}	
\nonumber
\end{split}
\eeq
so the magnitude covariance matrix for the {\it scaled} velocity power spectrum is given by 
\beq
\begin{split}
C^{m}_{i j }&(A_{1},A_{2}... A_{N}) = \\
& \left( \frac{5} {\ln{10} }\right)^{2} \left( 1 - \frac{(1 + z_{i})^{2}}{H(z_{i})d_{L}(z_{i})}\right) \left( 1 - \frac{(1 + z_{j})^{2}}{H(z_{j})d_{L}(z_{j})}\right) 	\\
& \sum_{i =1}^{N} A_{i} \int^{k^{\text{cen}}_{i} + \Delta_{i}/2}_{k^{ \text{cen} }_{i} -\Delta_{i}/2} \frac{d k}{ 2\pi^{2}} k^{2}{\cal P}_{ v v }^{\text{Fid}}(k) W(k,\alpha_{i,j},r_{i},r_{j}).	
\label{cov1new}		
\nonumber
\end{split}
\eeq
 From Eq.~(\ref{like}) and Eq.~(\ref{Sig}) we then have 
\beq
\begin{split}
P(A_{1},& A_{2},... A_{N} | {\bf \delta m}) =|2\pi \Sigma(A_{1},A_{2},... A_{N}) |^{-1/2} \\
&\exp{\left(-\frac{1}{2}{\bf \delta m}^{\text{T}} \Sigma^{-1}(A_{1},A_{2},... A_{N} ){\bf \delta m} \right)}.
\label{likenew}
\end{split}
\eeq
The best-fitting values $A_{i}$ can be used to check 
the consistency with the fiducial model $(A_{i} = 1)$
or to obtain the effective measured power ${\mathcal P}_{i}$ 
in each bin:
\beq
 {\mathcal P}_{i} = A_{i}\int^{k^{\text{cen}}_{i} + \Delta_{i}/2}_{k^{ \text{cen} }_{i} -\Delta_{i}/2} \ d k\frac{ {\cal P}_{ v v }(k) }{\Delta_{i}}.
\label{v_mean}
\eeq
The ${\mathcal P}_{i}$ values can now be compared with the 
predictions of the velocity power 
spectrum from different cosmological models.

\subsubsection{Scale-dependent growth rate \label{sec:growthpar}}

We can also relate the measured $A_{i}$ values to the growth 
rate of structure at each scale, as follows.

Here we will assume linear perturbation theory to be valid for both 
the density and the velocity fields; the justification 
for this assumption will be given in Section \ref{sec:gridding_met}.  
In this regime the linear continuity equation is valid i.e., $\theta(k) =- f H \delta(k)$. 
These assumptions are required to place constraints on the 
growth rate, but {\it not} required
for the previous parametrisations. 
A shift in $f(z)\sigma_{8}(z)$ from the fiducial value to a new value, viz., 
$ f\sigma_{8}(z)^{\rm Fid.} \rightarrow f\sigma_{8}(z) $, has an effect on the velocity divergence power spectrum
that can be calculated as ${\mathcal P}_{\theta \theta}(k) \rightarrow A_{1}{\mathcal P}_{\theta \theta}(k)$,
where $A_{1} =\left( f\sigma_{8}(z)/ f\sigma_{8}(z)^{\rm Fid} \right)^{2}$. 
One can then write down a `scaled' velocity divergence
power spectrum as
\beq
\begin{split}
& {\cal P}_{ \theta \theta}^{\text{Scaled}}(k) \equiv \\ 
&\left( f\sigma_{8}(z,k^{\text{cen}}_{1})/ f\sigma_{8}(z)^{{\rm Fid}} \right)^{2}{\cal P}_{ \theta \theta}^{\text{Fid}}(k)\Pi(k,\Delta_{1}, k^{\text{cen}}_{1})  \\
&+ \left( f\sigma_{8}(z,k^{\text{cen}}_{2})/ f\sigma_{8}(z)^{\rm Fid} \right)^{2}{\cal P}_{ \theta \theta}^{\text{Fid}}(k)\Pi(k,\Delta_{2}, k^{\text{cen}}_{1})  \\
& ... + \left( f\sigma_{8}(z,k^{\text{cen}}_{N})/ f\sigma_{8}(z)^{\rm Fid} \right)^{2}{\cal P}_{ \theta \theta}^{\text{Fid}}(k)\Pi(k,\Delta_{N}, k^{\text{cen}}_{N}),
\label{eqn:scaled_growth}
\end{split}
\eeq
where again  ${\mathcal P}_{v v}^{\text{Scaled}}(k) \equiv {\cal P}_{ \theta \theta }^{\text{Scaled}}(k)  /k^{2} $, and 
there are $N$ different bins that span the entire $k$ range.
The growth rate is considered to be constant over the 
wavenumber range of a given bin. The above relation Eq. (\ref{eqn:scaled_growth})
results from the approximation ${\mathcal P}_{\theta \theta}(k,z) \propto \left( \sigma_{8} f(k,z)\right)^{2}$.

The velocity power spectrum is calculated (at $z=0$) by assuming the standard 
$\Lambda$CDM expansion history and that the growth of perturbations is governed by GR.
We note that modifying the expansion history and/or deviations from GR at higher
redshifts will affect the current growth rate. 
Therefore in order to consistently 
examine the possibility of a scale-dependence 
of the growth rate of structure (i.e., moving beyond a consistency test) 
such effects would need to be taken into account.
Such an approach is beyond the scope of this 
paper and left for future work; here we simply consider if the observed growth 
rate as a function of scale is consistent with that expected within the 
framework of the standard model.

\subsection{Modelling of the velocity power spectrum \label{sec:vpower} }
In this section we will outline the model we use for the velocity power spectrum
in terms of the cosmological parameters. 

We calculate the real-space velocity power spectrum using the code {\tt velMPTbreeze} 
(an extension of {\tt MPTbreeze} in \citet{Crocce:2012kl}), 
which computes the velocity power spectrum using two loop multi-point propagators 
\citep{Bernardeau:2008fk} in a similar way to renormalized perturbation 
theory (RPT) \citep{Crocce:2006tg}. 
{\tt velMPTbreeze} uses an effective description of multi-point propagators
introduced in \citet{Crocce:2012kl}
which significantly reduces computation time relative to other RPT implementations. 
The results from {\tt velMPTbreeze} were extensively tested against $N$-body 
simulations (Crocce and Scoccimarro, in prep).

\subsection{Reducing non-linear systematics and computation time\label{sec:gridding_met}}

The velocity field is directly driven by the tidal gravitational field $\nabla \Phi$, where $\Phi$ is the 
gravitational potential, which causes it to depart from the linear regime at larger scales 
than the density field \citep{Scoccimarro:2004ij}. While the off-diagonal 
elements of the covariance matrix Eq.~(\ref{cov11}) are 
dominated by large-scale modes,
as a result of the survey geometry\footnote{This can be seen when plotting the window function 
$W(k) \equiv \left(\sum_{j=1}^{N} \sum_{i=1}^{N} W(k,\alpha_{i j},r_{i},r_{j})\right)/N^{2}$ 
of the survey (where $W(k,\alpha_{i j},r_{i},r_{j})$ is defined in Eq.~(\ref{window})) and
$N$ is the number of galaxies in the survey. This window function only influences off-diagonal 
elements of the covariance matrix. One finds that
the amplitude of $W(k)$ significantly reduces as small-scales are approached, therefore less weight is attached to the 
power spectrum on small scales.},
this is not the case for the 
diagonal (cosmic variance) elements where the small scale power contributes to 
the intrinsic scatter. Hence non-linear effects are important to consider and 
minimize.

In order to suppress non-linear contributions and hence
reduce potential systematic biases we adopt 
a simple smoothing (gridding) procedure. Gridding the velocity field significantly reduces
the computation time by reducing the size of the covariance matrix; this will
be essential for next-generation data sets given the computational 
demands of the likelihood calculation (which requires a matrix inversion for each 
likelihood evaluation).

\begin{figure}
\includegraphics[width=8.5cm]{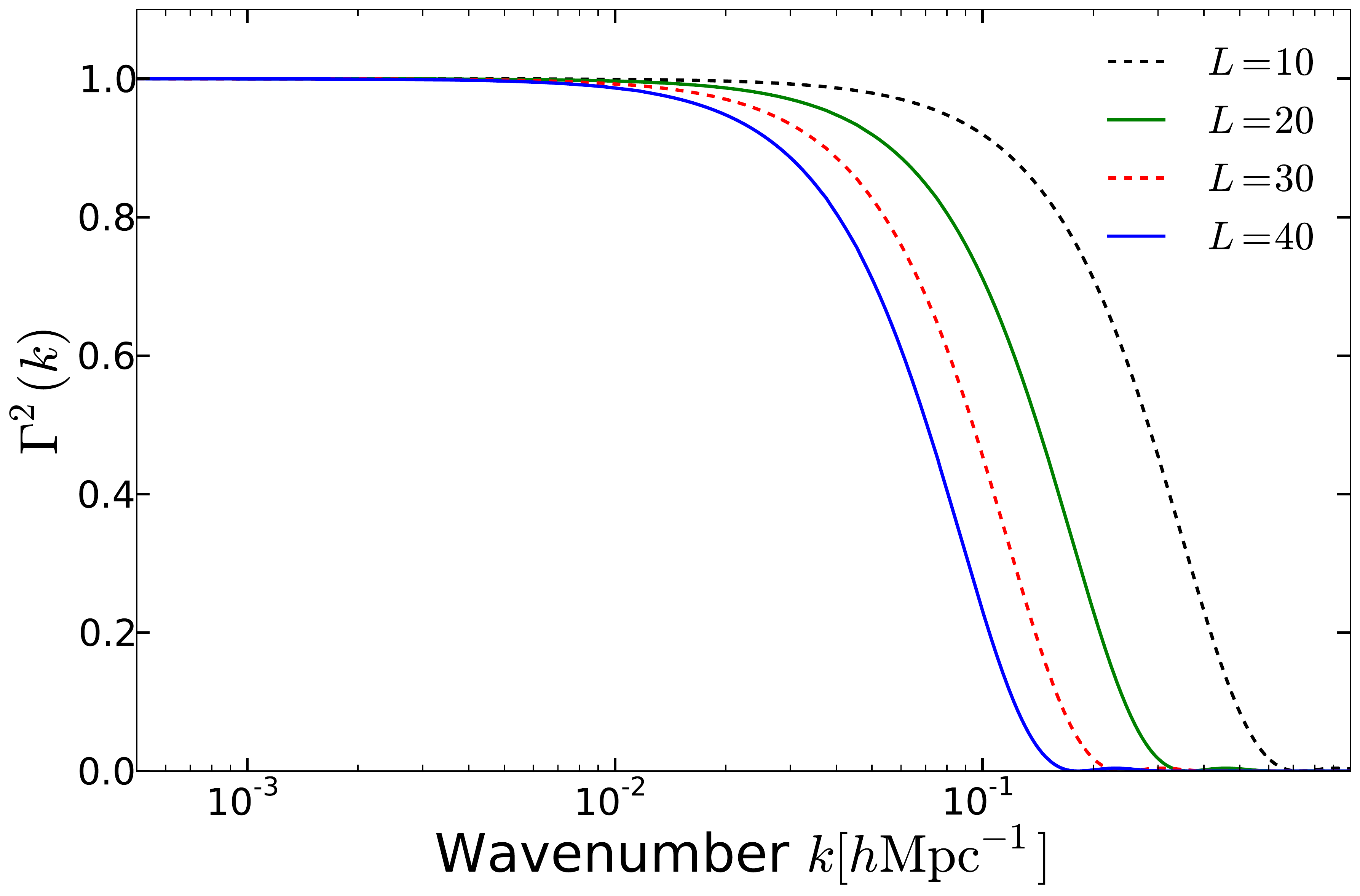}
\caption{Examples of the smoothing kernel $ \Gamma(k) \equiv {\mathcal F}[\Theta_{i j}]$ 
for different values of the smoothing length $L$, given in units of $h^{-1}$~Mpc. We plot the square of the kernel as this is the term
that modulates the velocity power spectrum, i.e., the term occuring in eq (\ref{cov11}).}
\label{fig:gamma_fn}
\end{figure}

The binning method we implement was developed 
and tested in \citet{Abate:2008bs}. 
The grid geometry used is a cube of length $L$, where the average apparent magnitude fluctuation 
$\delta m$ and error $\sigma_{\delta m}$
are evaluated at the centre of the $i^{\text{th}}$ grid cell $\vec{x}_{i}$:
\beq
\begin{split}
\delta m_{i}(\vec{x}_{i} ) &= \frac{1}{N_{i}}\sum_{j}\delta m_{j}^{\hspace{0.2mm}\text{gal}}(\vec{x}_{j})\Theta_{i j},\\
\sigma_{\delta m, i} &= \frac{1}{N_{i}^{3/2} }\sum_{j}\sigma_{\delta m, j}^{\hspace{0.2mm}\text{gal}}\Theta_{i j},
\label{eqn:gridnew}
\end{split}
\eeq 
where $N_{i}$ is the number of galaxies located within the $i^{\text{th}}$ cell,
$\delta m^{ \text{gal}}$ is the inferred fluctuation in apparent magnitude for a specific galaxy and
$\sigma_{j}^{\hspace{0.2mm}\text{gal}}$ is the error component as defined in Eq.~(\ref{err}). 
The optimal choice for the gridding length scale is evaluated using numerical simulations and 
is discussed in Section \ref{sec:sim}.
Both the observational error from the distance indicators and the error introduced by the
non-linear velocity dispersion $\sigma_{\rm v}$ are being averaged. 
The sum over $j$ is taken over the entire sample, where $\Theta_{i j}$ equals one when the galaxy 
is within the grid cell and zero otherwise.
The process of smoothing the velocity field effectively damps
the velocity power spectrum, this acts to suppress non-linear contributions.
The function describing this damping is given by the Fourier transform of the kernel
$\Theta_{i j}$, introduced in Eq.~(\ref{eqn:gridnew}). 
Letting $ \Gamma(k) \equiv {\mathcal F}[\Theta_{i j}] $ from above we have
\beq
\Gamma(k) = \left\langle \text{sinc}\left(k_{x}\frac{L}{2} \right)\text{sinc}\left(k_{y}\frac{L}{2} \right)\text{sinc}\left(k_{z}\frac{L}{2} \right)\right\rangle_{\vec{k} \in k},
\eeq
{where $\langle F(\vec{k}) \rangle_{\vec{k} \in k}$ is the expectation value of $F(\vec{k})$ in the phase space $\vec{k} \in k$ i.e., $\langle F(\vec{k})\rangle_{\vec{k} \in k} = 1/4\pi \int d\Omega F(\vec{k})$. Examples of $\Gamma(k)^{2}$, for a range of different smoothing scales, are given in Fig. \ref{fig:gamma_fn}}.
This allows one to calculate the velocity power spectrum between separate grid points; therefore 
once the velocity field has been smoothed we alter the theoretical prediction of the velocity power spectrum 
by ${\mathcal P}_{v v}(k) \rightarrow {\mathcal P}^{\text{Grid}}_{v v}(k) ={\mathcal P}_{v v}(k)\Gamma^{2}(k).$
Now the covariance of $\delta m$ between grid centres,
$\tilde{C}_{i j}$, is given by
\begin{equation}
\begin{split}
 \tilde{C}_{i j } =& \left( \frac{5} {\ln{10} }\right)^{2} \left( 1 - \frac{(1 + z_{i})^{2}}{H(z_{i})d_{\rm L}(z_{i})}\right)	\left( 1 - \frac{(1 + z_{j})^{2}}{H(z_{j})d_{\rm L}(z_{j})}\right) 	\\
&\int \frac{d k}{ 2\pi^{2}} k^{2}{\cal P}_{ v v}(k,a=1)W(k,\alpha_{ij},r_{i},r_{j})\Gamma^{2}(k).
\label{cov11}		
\end{split}
\end{equation}

Using numerical $N$-body simulations \citet{Abate:2008bs} explore
the dependence of the recovered best-fitting parameters ($\sigma_{8}$ and $\Omega_{\rm m}$)
on the smoothing length. 
Specifically they find that 
(relative to the statistical error) a smoothing 
scale greater than $10h^{-1}$Mpc results in an unbiased estimation of the cosmological 
parameters of interest.

In order to derive Eq. (\ref{cov11}) one must presuppose the PVs
inside each cell are well-described as a continuous field. However the velocities 
inside a grid cell represent discrete samples from the PV field; therefore as the 
number density inside each cell becomes small this approximation becomes worse.
In \citet{Abate:2008bs} a solution to this `sampling problem' was proposed and tested 
using $N$-body simulations. 
To mitigate the effects of this approximation one interpolates
between the case of a discrete sample and that of the continuous field limit. 
The weight attached to each is determined using the number of 
galaxies within each cell $N_{i}$. The diagonal elements of the covariance matrix are
now updated as
\beq
\tilde{C}_{i i }  \rightarrow \tilde{C}_{i i} +
(\tilde{C}_{i i } - C_{i i}^{\rm m} )/N_{i}, 
\label{updatecov}
\eeq
where $C_{i i}^{\rm m}$ is defined in Eq. (\ref{cov11old}).
For this correction the continuous field approximation is assumed
for the off-diagonal elements\footnote{This approach is valid given the 
off-diagonal elements of the covariance matrix are significantly damped at 
small scales, and hence 
the smoothing of the velocity field has only a small effect on these 
elements.}.

\subsection{Effect of the unknown zero-point\label{sec:zeropoint}}

The zero-point in a PV analysis is a reference magnitude,
or size in the case of Fundamental Plane surveys,  for which
the velocity is known to be zero. From this reference point one is able to 
infer the velocities of objects; without such a reference point only the relative velocities 
could be determined. 
An incorrectly calibrated zero-point introduces a monopole component 
to measured PVs.
To give an example,
for supernovae the zero-point is determined by 
the absolute magnitude $M$ and the Hubble parameter $H_{0}$.

When deriving PV measurements 
the zero-point is typically fixed at its maximum likelihood value 
found during the calibration phase of the analysis;
this allows the 
velocities of all the objects in the sample to be determined.
However, this zero-point may contain error.
In this section we introduce a method to analytically
propagate the uncertainty in the zero-point
into the final cosmological result. 

We first consider the case of analysing a single velocity survey.
We define $a$ as an offset in the magnitude fluctuation;
such that $\delta m \rightarrow \delta m + a$. This
indirectly represents a perturbation to the velocity zero-point.
Given we have some prior knowledge of the distribution of this variable 
we give it a Gaussian prior i.e., 
\beq
P(y|\sigma_{y}) = \frac{1}{(2\pi)^{1/2}\sigma_{y}}\exp[-y^{2}/2\sigma^{2}_{y}].
\label{gauss:prior}
\eeq
We define ${ \bf x}$ as an $N$ dimensional vector where each element is 
set to one (i.e., $({ \bf x})_{i} = 1$, for $i = 1..N$). 
Here $N$ is the dimension of ${\bf \delta m}$.
The parameter $a$ alters the theoretical prediction
for the mean velocity, $\langle{\bf \delta m}^{p}\rangle = 0 $, to $\langle{\bf \delta m}^{p}\rangle =y {\bf x}$. Now we 
can analytically marginalize over the unknown zero-point \citep{S.L.Bridle:2002fv}
\beq
\begin{split}
& P(\Sigma |\delta {\bf m} ) = \int dy \hspace{1mm}P( \Sigma |\delta {\bf m},y)P(y|\sigma_{y}) \\
&= |2\pi\Sigma|^{-1/2}(1 + {\bf x }^{T}\Sigma^{-1}{\bf x }\sigma_{y}^{2})^{-1/2} \exp\left[\frac{1}{2}{ \bf \delta m}^{T} \Sigma_{M}^{-1} {\bf  \delta m} \right],
\end{split}
\label{eqn:zero1}
\eeq
where
\beq
\Sigma_{M}^{-1} \equiv \Sigma^{-1} - \frac{\Sigma^{-1}{\bf x }{\bf x}^{\text{T}}\Sigma^{-1}}{ {\bf x }^{\text{T}}\Sigma^{-1}{\bf x } + \sigma_{y}^{-2}} .
\eeq
We may wish to combine a number of different PV surveys 
with potentially different 
zero-point offsets. In this case it is necessary to consider how 
one can marginalise over the independent zero-points simultaneously.
We consider the example of 
two different PV surveys but note that this approach can be readily generalised to a larger number 
of surveys \citep{S.L.Bridle:2002fv}. 

Firstly we decompose the data vector into apparent magnitude fluctuations from 
the first and second surveys,
\beq
\vec {\delta m}  = 
\left(
\begin{array}{c}
\vec {\delta m}^{(1)}\\
\vec {\delta m}^{(2)}\\
\end{array}
\right)_{N},
\eeq
where the first survey has $n_{1}$ data points and the second has $n_{2}$, therefore the combined vector has length $N = n_{1} + n_{2}$.
The data from the two surveys needs to be smoothed onto two different grids, this is a simple modification to the binning algorithm:
\beq
\vec {\delta m}  = 
\left(
\begin{array}{c}
 \frac{1}{N_{1, i}}\sum_{j \leq n_{1}}\delta m_{j}^{\hspace{0.2mm}\text{gal}}(\vec{x}_{j})\Theta_{i j}\\
  \frac{1}{N_{2,i}}\sum_{n_{1} < j \leq n_{2}}\delta m_{j}^{\hspace{0.2mm}\text{gal}}(\vec{x}_{j})\Theta_{i j}\\
\end{array}
\right),
\label{sep_grid}
\eeq
where $N_{1, i} $ and $N_{2, i}$ are the number of galaxies inside the i$ ^{th} $ cell from the first and second 
survey respectively.

We now introduce two free parameters $(y,b)$ which will allow the zero-point to vary for each survey,
again both parameters are given Gaussian priors (i.e., are distributed according to Eq.~(\ref{gauss:prior})).
To account for a changing zero-point we alter the theoretical prediction for the mean value of the apparent magnitude fluctuations  $\langle \bf {\delta m}^{p}\rangle$.
This quantity is normally set to zero as PVs are assumed to be distributed according to a multivariate Gaussian with a mean of
zero, now we have $\langle {\bf \delta m}^{p} \rangle =y{\bf x}^{(1)} + b {\bf x}^{(2)}$ where ${x}^{(1)}_{i} = 1 $ if $ i \leq  n_{1}$ and ${x}^{(1)}_{i} =0$ otherwise 
and ${x}^{(2)}_{i} = 1 $ if $ i \geq  n_{1}$ and ${x}^{(2)}_{i} =0$ otherwise. The 
updated likelihood is then
\beq
\begin{split}
&P( \Sigma |\delta {\bf m},y,b )=\\
&|2\pi \Sigma |^{-1/2}\exp{\left(-\frac{1}{2}\left({\bf \delta m} +\langle \bf {\delta m}^{p}\rangle \right)^{\text{T}} \Sigma^{-1}\left({\bf \delta m} + \langle \bf {\delta m}^{p}\rangle\right) \right)}.
\end{split}
\nonumber
\eeq
We desire a posterior distribution independent of the zero-point corrections therefore 
we analytically marginalise over these parameters
\beq
\begin{split}
P(\Sigma |\delta {\bf m} ) & = \int dy\int db \hspace{1mm} P( \Sigma |\delta {\bf m},y,b )P(y| \sigma_{y})P(b| \sigma_{b}) \\
&= |2\pi\Sigma|^{-1/2}(1 + {\bf x^{(1)} }^{T}\Sigma^{-1}{\bf x^{(1)} }\sigma_{y}^{2})^{-1/2} \\
&(1 + {\bf x^{ (2)} }^{T}\Sigma^{-1}{\bf x^{(2)}}\sigma_{b}^{2})^{-1/2} 
\exp\left[\frac{1}{2}{ \bf \delta m}^{T} \Sigma_{M}^{-1} {\bf  \delta m} \right],
\end{split}
\label{eqn:zero2}
\eeq
where
\beq
\Sigma_{M}^{-1} \equiv \Sigma^{-1} - \frac{\Sigma^{-1}{\bf x^{(1)} }{\bf x^{(1)}}^{\text{T}}\Sigma^{-1}}{ {\bf x^{(1)} }^{\text{T}}\Sigma^{-1}{\bf x^{(1)} }}
- \frac{\Sigma^{-1}{\bf x^{(2)} }{\bf x^{(2)}}^{\text{T}}\Sigma^{-1} + \sigma_{y}^{-2}}{ {\bf x^{(2)} }^{\text{T}} \Sigma^{-1}{\bf x^{(2)} } + \sigma_{b}^{-2}}.
\eeq
Here we need to consider the variation to 
the determinant as 
 the covariance matrix is being varied at each likelihood evaluation.
For all zero-points here we choose a Gaussian prior with a standard deviation of 
$\sigma_{y}= \sigma_{b} = 0.2$. We find the choice of width of the prior has 
an insignificant effect on the final results. 

\subsection{Combining multiple (correlated) velocity surveys\label{sec:combsurveys}}

Given the limited number count and sky coverage of objects in velocity surveys it is
common for different surveys to be combined in a joint analysis. 
In this situation individual datasets may contain unrecognised systematic
errors, requiring them to be re-weighted in the likelihood analysis.

The first method we consider to do this is a recent upgrade to the hyper-parameter
analysis. The original hyper-parameter method
was developed
to remove the inherent subjectivity associated with selecting which data 
sets to combine in an analysis and which to exclude  \citep[see][]{Lahav:2000dz, Hobson:2002fu}. 
This process is achieved
by including all the available data sets but allowing free hyper-parameters to vary
the relative `weight' attached to each data set, the hyper-parameters
are then determined in a Bayesian way.  Consider two hypothetical 
surveys with chi squared of $\chi^{2}_{A}$ and $\chi^{2}_{B}$. 
The combined 
constraints are typically found by minimising the quantity
\beq
\chi^{2}_{\text{com}} = \chi^{2}_{\rm A} + \chi^{2}_{\rm B}.
\label{hp_1}
\eeq
This gives both data sets equal weight. Introducing the
hyper-parameters one has
\beq
\chi^{2}_{\text{com}} = \alpha \chi^{2}_{\rm A} + \beta \chi^{2}_{\rm B}.
\label{hp_2}
\eeq
The hyper-parameters can be interpreted as scaling the errors for each data 
set, i.e., $\sigma^{i} \rightarrow$ $\sigma_{i} \alpha^{-1/2}$, or 
equivalently the covariance matrix of each data set $C_{i} \rightarrow \alpha^{-1} C_{i}$.
The final values of the hyper-parameters,
more accurately their probability distributions $P(\alpha)$ and $P(\beta)$, 
give an objective way to determine if 
there are systematic effects present in the data (e.g., a value $\alpha > 1$ can
be interpreted as reducing the errors or correspondingly increasing the 
relative weight of the data set).

The problem with 
the traditional hyper-parameter analysis for PV surveys is
that it assumes that the individual data sets are {\it not correlated} (this 
assumption is required to write down equation Eq.~(\ref{hp_1}) and Eq.~(\ref{hp_2})).
If the surveys cover overlapping volumes or are influenced by the same 
large-scale modes this is not the case. 
Recently the hyper-parameter formalism has been extended to
a hyper-parameter matrix method which includes
the cross correlations between surveys \citep{Ma:2013kl}. 
Here the hyper-parameters scale both the covariance between 
objects in a given data set and the covariance between the data sets:
\beq
C^{D_{i} D_{j}} \rightarrow \left(\alpha_{i} \alpha_{j} \right)^{-1/2} C^{D_{i} D_{j}} 
\eeq
$D_{i}$ represents the i$^{\text{th}}$ data set, so $C^{D_{i} D_{j}}$ gives the covariance 
between the i$^{\text{th}}$ and j$^{\text{th}}$ data sets. 
For simplicity here we outline the case of two different data sets.
In this case there are two hyper-parameters $(\alpha_{1},\alpha_{2})$ which we 
treat as free parameters.  
The hyper-parameter matrix is defined as:
\begin{equation}
H=\left(
\begin{array}{cc}
\alpha _{1}^{-1} & (\alpha _{1}\alpha _{2})^{-1/2}  \\
(\alpha _{1}\alpha _{2})^{-1/2} & \alpha _{2}^{-1}  \\
\end{array} %
\right) .  \label{hyper_matrix}
\end{equation} %

The final likelihood function is
\beq
\begin{split}
&P({\bf \delta m}|\vec{\theta},\vec{\alpha})=\\
 &\left[
\prod\limits_{i=1}^{2}\left(\frac{\alpha_i}{2\pi}\right)^{n_{i}/2}\right]
 \frac{1}{\sqrt{ |C|}}\exp \left( -\frac{1}{2}%
\delta m^{T}\left( \hat{H}\odot C^{-1}\right)
\delta m\right)\,.
\label{like4-copy}
\end{split}
\nonumber
\eeq
Here $\odot$ is an `element-wise' product (or, Hadamard product) defined as 
$(\hat{H}\odot C^{-1})_{i j } =\hat{H}_{i j}\times (C^{-1})_{i j}$, and $\vec{\theta}$
represents the parameters of interest. 
$\hat{H}$ is the Hadamard inverse of the `hyper-parameter' matrix
(i.e. $\hat{H}_{i j} = P_{i j}^{-1}$), and $n_{1}$ and $n_{2}$ 
are the number of data points in the first and second surveys respectively.

As described in Section \ref{test_like} a free parameter 
$\sigma_{\rm v}$ is typically introduced to account for non-linear random
motion. One issue with the likelihood function defined above is that 
$\sigma_{\rm v}$ and the hyper-parameters are quite degenerate.
Therefore for our hyper-parameter analysis we fix $\sigma_{\rm v}$
at the values found when analysing the surveys independently.

\section{Testing with simulations}
\label{sec:sim}

We require simulations of PV catalogues for several aspects of this analysis. Firstly,
to determine if non-linear effects from the growth rate of structure or redshift-space distortions
cause systematic errors.
Secondly, to determine the approximate survey geometry and distance errors for which the 
non-Gaussian observational scatter of PVs becomes important. Finally to determine
the effect (on the final constraints) of marginalising over the zero-point uncertainty.
Note the construction of the mock catalogues used in this section is outlined in 
Section \ref{sec1}.

All the 
cosmological parameters not allowed to vary freely here are set to those input into 
the simulation (i.e., $\Omega_{\Lambda} = 0.727, \Omega_{\rm m} = 0.273, \Omega_{k} =0, H_{0} = 100h~{\rm km~{s}^{-1}}{\rm {Mpc}^{-1}},
\sigma_{8} = 0.812, n_{\rm s} = 0.960$). For the velocity power spectrum fits 
we use a smoothing scale (defined in Section \ref{sec:gridding_met}) of $10h^{-1}$Mpc, while for 
the analysis of $\Omega_{\rm m}$ and $\sigma_{8}$ we adopt a length of $20h^{-1}$Mpc.
We use a larger grid size for the analysis of $\Omega_{\rm m}$ and $\sigma_{8}$ 
because the evaluation of the likelihood (i.e., Eq.~\ref{eqn:likom}) is more computationally 
demanding relative to the evaluation of of the likelihood given in Eq.~(\ref{likenew}), the 
larger grid size reduces the computational requirements\footnote{
This is the case because for each $\Omega_{\rm m}$ and $\sigma_{8}$ posterior evaluation we are 
required to re-calculate the entire covariance matrix (Eq.~\ref{cov11}). 
This is not the case for the other parametrisations considered here.}.
We first shift the haloes within the 
simulation to their redshift-space position, using
${ \bf x}^{s} = { \bf x}^{r} + {\bf v}({\bf x},t) \cdot \hat{r}/H_{0}$. 
Now we transform the 
PVs within the simulation to apparent magnitude fluctuations, $\delta m$. 

At small scales the predictions from RPT become less accurate and are known to 
break down (experience exponential damping relative to the expectations from $N$-body simulations) 
at $k\sim 0.15 h{\rm Mpc^{-1}}$
for the velocity power spectrum evaluated assuming the fiducial cosmology of the 
simulation at a redshift of zero.
We therefore truncate the velocity power spectrum fits at this scale. We note that this scale varies for 
different cosmological parameters, therefore for the ($\Omega_{\rm m},\sigma_{8}$) fits we
test a range of values, $k_{\text{max} }$, for truncating the integral 
when calculating the covariance matrix, to decide the optimal choice for the 
data.

Now using 8 different observers from 
mock set (\rom{1}) we test the ability of each parametrisation to recover the 
input cosmology, under the conditions outlined above. 
Recall for mock set (\rom{1}) the input distance error is
$\sigma_{\rm d} \sim 5\%$, the approximate distance error for SNe. 
The derived constraints on ($\Omega_{\rm m},\sigma_{8}$) for various values of $k_{\text{max}}$ are given 
in Fig. \ref{check_code:1}; the black square symbols here give the input cosmology of the simulation.
The velocity power spectrum measurements are given in 
Fig. \ref{check_code:2} and the constraints for a scale-dependent growth rate, $f\sigma_{8}(z=0, k)$,
are given in Fig. \ref{check_code:3}. The thick blue lines in Fig. \ref{check_code:2}
give the predictions 
for the average power within the defined bin ranges for the fiducial cosmology,
this is calculated using Eq.~(\ref{v_mean}) with $A_{i} =1$. 
In addition to giving the results for a single mock realisation
we also average the results found for 8 different mock realisations in order to provide a more 
accurate systematic test. 
Again some care needs to be taken 
when interpreting the combined constraints given that on the largest scales the mock
realisations are {\it significantly correlated}. 
This is most pronounced for the largest-scale bin in Fig. \ref{check_code:2}  and Fig. \ref{check_code:3},
for which we interpret the consistently `high' measurement power as being produced by correlations.
Also note 
the mock simulations considered here have {\it significantly} greater statistical power than current
PV surveys, so we are performing a sensitive systematic check. We find that at the investigated error levels 
we are able to accurately 
recover the input cosmology of the simulation for all parametrisations considered.
We conclude therefore that the bias from non-linear structure is currently insignificant, the linear relation 
between the PV and $\delta m$ is valid and non-linear RSD effects do not bias our final constraints.

\begin{figure}
\centering
 \includegraphics[width=9cm]{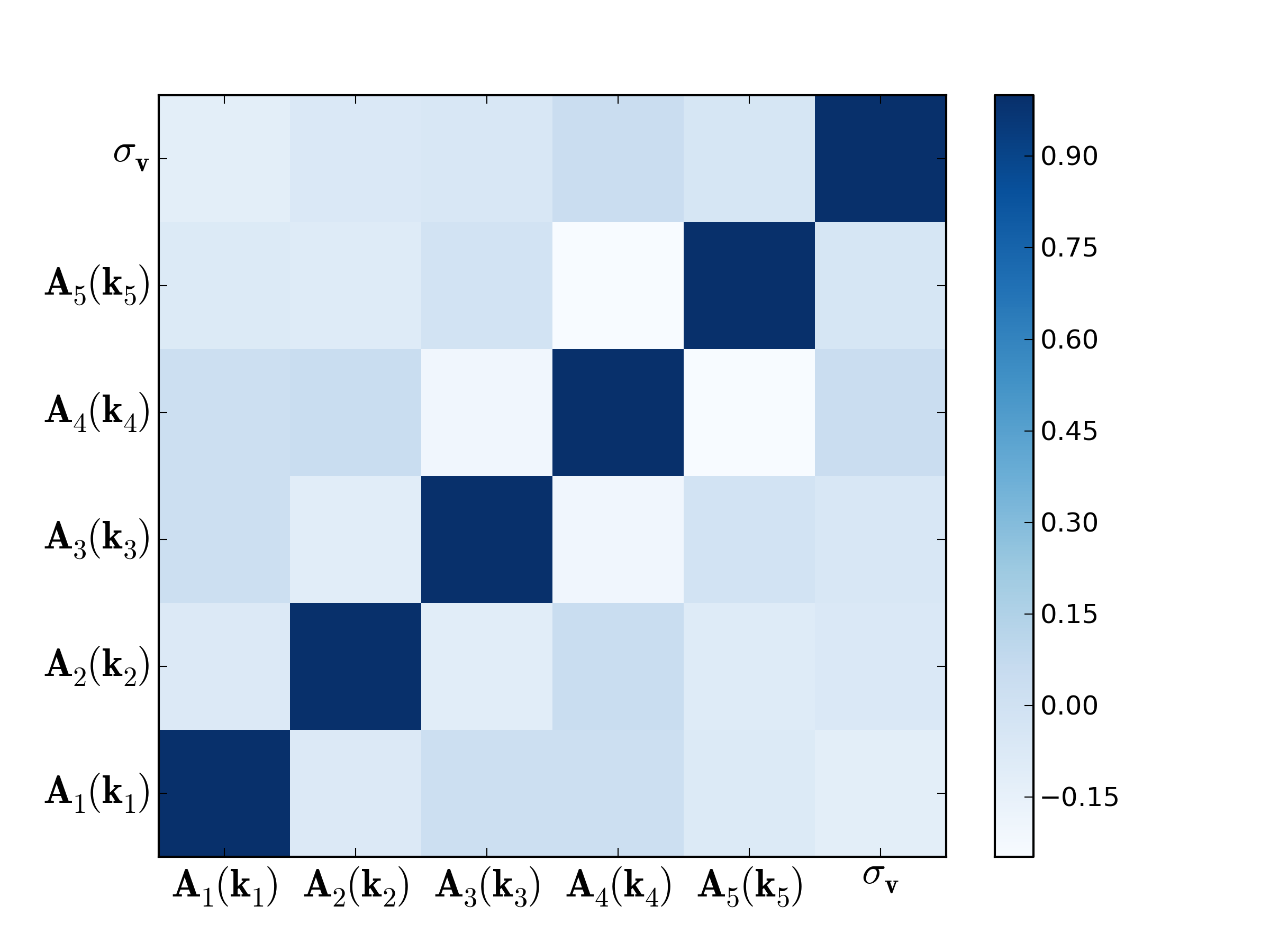}
\caption{Correlation coefficents $r$ between the amplitude parameters $A_{i}$, and the non-linear
velocity dispersion $\sigma_{\rm v}$. The results here were calculated using an MCMC chain
(of length $\sim 10^{6}$) produced when analysing a single realisation from Mock set (\rom{1}).
We expect very similar correlations to exist between the growth rate measurements and 
note that the correlations between the different bins are quite weak.}
\label{fig:correlation_mcmc}
\end{figure}

\begin{figure*}
\centering
 \includegraphics[width=18cm]{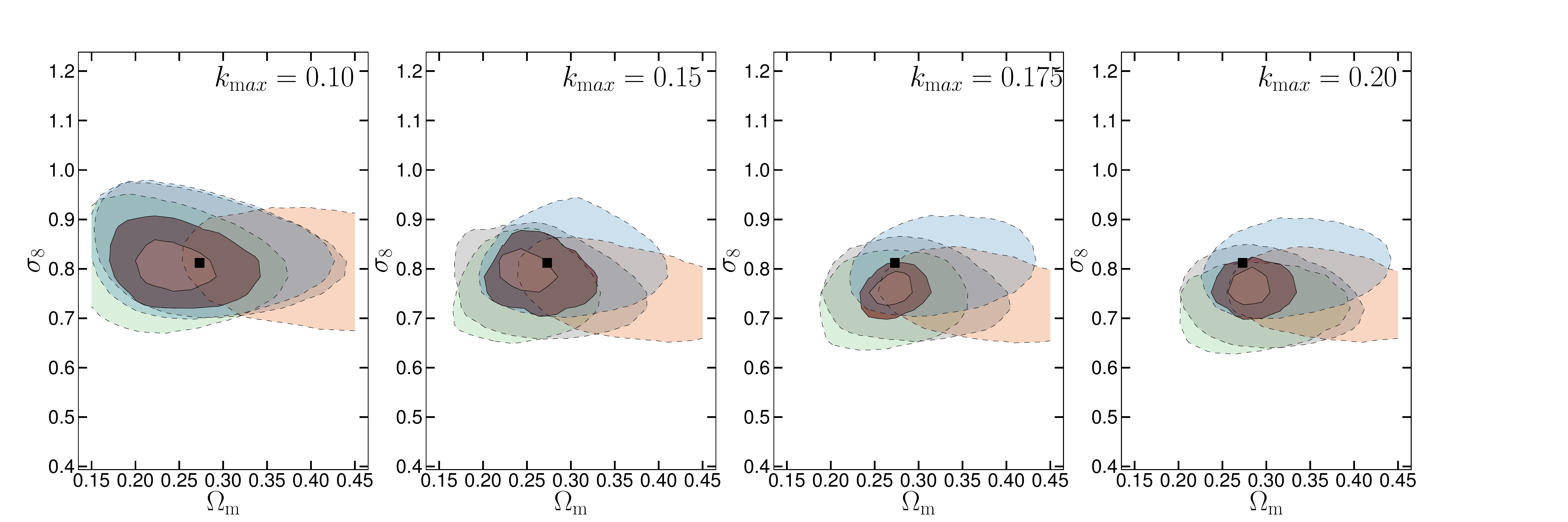}
\caption{ 68\% confidence regions for the matter density, $\Omega_{\rm m}$, 
and the RMS clustering in $8h^{-1}$Mpc spheres, $\sigma_{8}$, using mock set (\rom{1}),
including RSD and using the $\delta m$ variable. 
The transparent contours (dashed outline) give the constraints from some example single survey realisations. 
The opaque contours (solid outline) give the combined constraints from 8 realisations.
For the combined constraints we give 68\%  and 95\% confidence regions. 
A smoothing 
length of $20h^{-1}$Mpc is used for all constraints.
For each plot we vary the length scale, $k_{\text{max}}$ at which we truncate the integral for
the calculation of the covariance matrix, that is the integral given in Eq.~(\ref{cov11}) 
(i.e., the smallest scales included in the analysis). Varying this scale
allows us to test the validity of the constraints as we move into the non-linear regime. From 
left to right the wavenumbers at which we cut off the integration are $k_{\text{max}} = [0.1,0.15,0.175,0.20]h$Mpc$^{-1}$.
The black square symbols give the cosmology input into the simulation.} 
\label{check_code:1}
\end{figure*}

\begin{figure*}
\centering
\includegraphics[width=16cm]{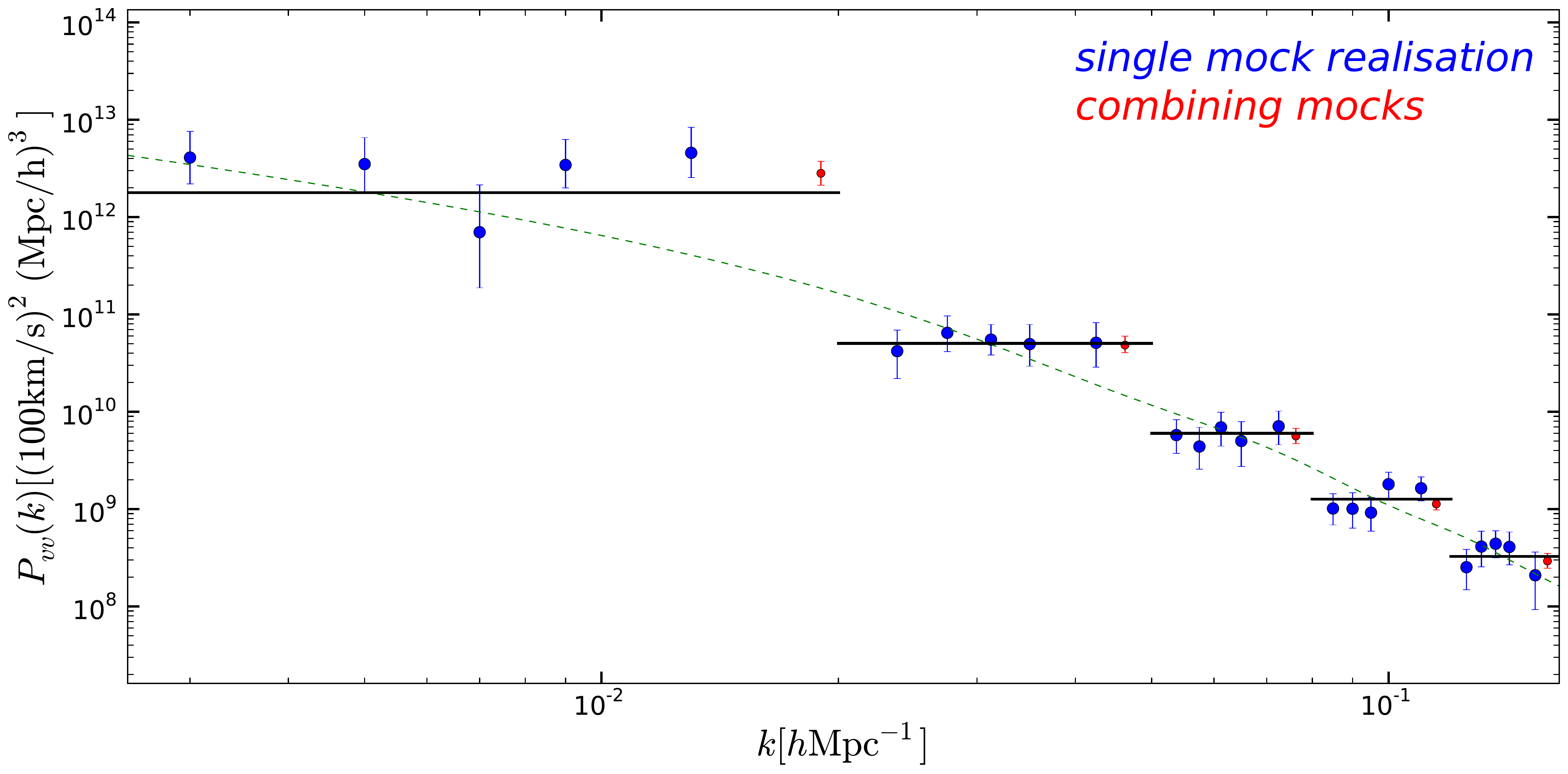}
\caption{ 68\% confidence intervals for the amplitude parameters $A_{i}$ describing the mean `power' within each bin
using mock set (\rom{1}). The thick blue (horizontal) lines give the mean power in each bin
for the fiducial cosmology calculated using Eq.~(\ref{v_mean}).
Here we include RSDs, use $\delta m$ and a smoothing 
length of $10 h^{-1}\rm{Mpc}$. The blue points are the constraints found 
for individual mock realisations, while the red points show the constraints found by combining the results from 8 different 
mocks. Consistency with the assumed fiducial cosmology occurs when the given confidence levels overlap with the 
mean power; the specific position of the point along the bin length is arbitrary. 
The green dashed line shows the velocity power spectrum calculated assuming the fiducial cosmology. 
Section \ref{par:2} gives the wavenumber bin intervals used here, with the exception that  $k_{\text{min}} = 0.0065h{\rm Mpc^{-1}} = 2 \pi/L_{\text{box}}$ .}
\label{check_code:2}
\end{figure*}

\begin{figure*}
\centering
\includegraphics[width=14cm]{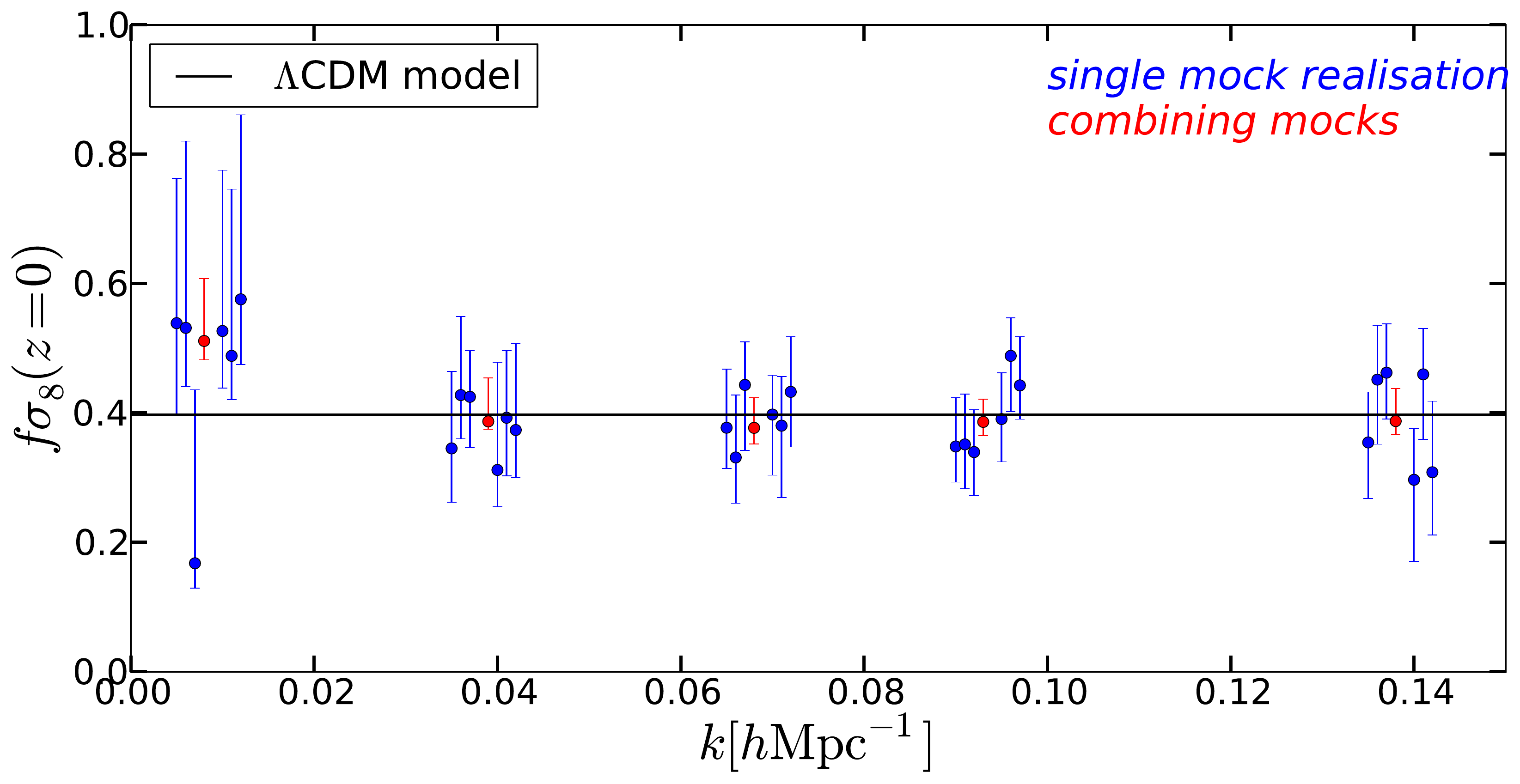}
\caption{68\% confidence intervals for the normalised {\it scale-dependent} growth rate $f(z=0,k)\sigma_{8}(z=0)$
in 5 different bins in Fourier space.
The thick black line gives the prediction of the input cosmology. For each $k$-bin we plot 
the results from 6 different realizations from mock set (\rom{1}). We include RSDs in the mocks, use the variable
$\delta m$, and choose a smoothing length of $10h^{-1}\rm{Mpc}$. 
The specific $k$ values within a given bin for the measurements are arbitrary. 
The bin intervals used here are given in Section \ref{par:2},
with the one correction that $k_{\text{min}} = 0.0065h^{-1}\rm{Mpc}$, 
corresponding to the size of the simulation.}
\label{check_code:3}
\end{figure*}

Following Fig. \ref{check_code:2} we conservatively fix $k_{\text{max}} = 0.15 h{\rm Mpc^{-1}}$ for the ($\Omega_{\rm m},\sigma_{8}$) fits,
given that on smaller scales we observe a slight trend away from the fiducial cosmology (yet still consistent at the $2\sigma$ level).
For the power spectrum fits we note a small amount of correlation exists between the different wavenumber bins. We give a
typical example of the correlation coefficients between the bins in Fig. \ref{fig:correlation_mcmc}, determined using the Monte 
Carlo Markov Chain.

\begin{figure*}
\centering
 \includegraphics[width=17cm]{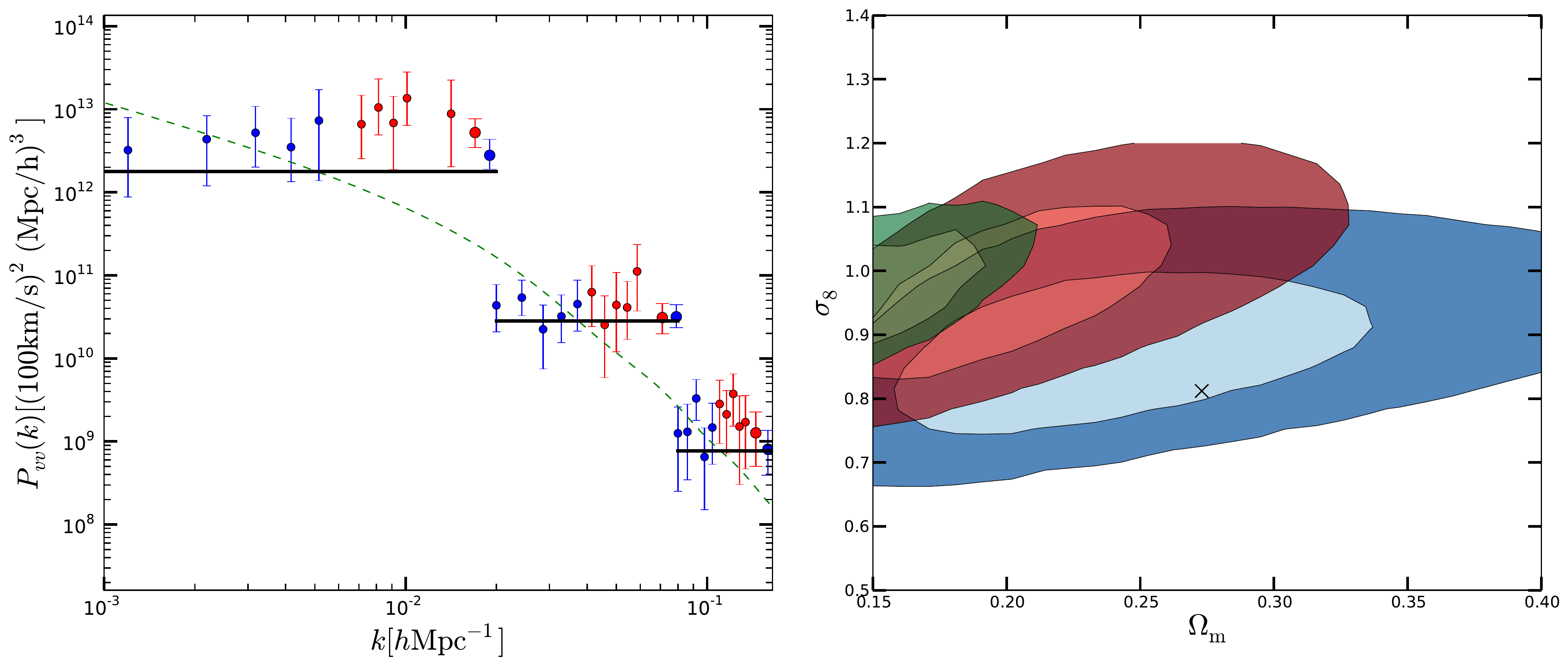}
\caption{(Left) $68\%$ confidence intervals for the velocity power spectrum amplitude in three Fourier bins. We consider five separate 
realisations taken from mock set (\rom{2}). The small blue points show the individual constraints found using the variable $\delta m$,
while the small red points show the constraints found using the median of the velocity distributions (viz, $v_{i} = \text{Median}[P(v_{i})]$)
(this gives very similar results to the direct method). 
The larger blue and red points show the results from combining the five realisations. The circle symbols (left panel)
give the median value of the probability distributions.
(Right) Constraints on the parameters $\Omega_{\rm m}$ and $\sigma_{8}$ found from combining the 
results from 8 different realisations with mock set (\rom{2}). 
The contours give $68\%$ and $95\%$ confidence
levels. The blue contour shows the result of using the variable $\delta m$.
The red and green contours show the result of 
using the PV as the main variable, where the red contour gives the result from
directly calculating the PV from the observable quantity
ignoring the  Jacobian term, and the green contour gives the constraints from using the mean value 
of $P(v)$.}
\label{check_code:4}
\end{figure*}

When testing the effect of non-Gaussian observational error for PVs, 
both the sky coverage of the survey and the distance error are relevant,
therefore we consider both mock set (\rom{1}) and (\rom{2}). 
We find that for mock set (\rom{1}) using the velocity not
magnitude as the 
variable in the analysis results in no significant bias. 
This continues to be true even when we limit the survey to one hemisphere. 
This can be understood because the degree of departure from Gaussianity of the 
probability distribution of peculiar velocities, $P(v)$, is 
dependent on the magnitude of the distance error.
With relatively small distance errors, $P(v_{p})$ is described well 
by a Gaussian distribution.

In the case of a distance and sky distribution corresponding to 6dFGSv, that is, 
$\sigma_{\rm d} \sim 30\%$ and only considering one hemisphere (i.e., mock set (\rom{2}))
we find a {\it significant} bias is introduced when using PVs\footnote{This also applies
for future analyses; a number of Fundamental Plane and 
Tully-Fisher surveys are forthcoming and will have similar properties.}. 
We use 8 realisations from mock set (\rom{2}),
generate realistic observational errors and perform the likelihood analysis
twice using either PV or $\delta m$ as the variable.
For the likelihood analysis using PV one is 
required to input a single velocity value, which gives us some freedom in how we 
choose to compress the distribution $P(v_{p})$ into a single value. 
Here we consider the mean, maximum likelihood (ML) and median.
For a detailed investigation into the effect of these choices, in the 
context of bulk flow measurements, see Scrimgeour et. al (in prep).
In all prior PV analysis when the full probability distribution 
of the distance measure (e.g., the absolute magnitude, $M$, in the case 
of the Tully--Fisher relation) was not available the 
PV was calculated directly from this variable. The Jacobian term
is ignored in this case, we label this method the `direct approach'.
To give an example; for the Fundamental Plane relation using this direct method one would determine the ML
value of  $x \equiv \log_{10}(D_{\text{z}}/D_{\text{H}})$ then using this value calculate the corresponding PV,
again ignoring the Jacobian term given in Eq.~(\ref{eqn:probcon}).

We give the constraints for the amplitude of the velocity power spectrum
and the cosmological parameters $\sigma_{8}$ and $\Omega_{\rm m}$,
found when using the magnitude fluctuation $\delta m$, in Fig. \ref{check_code:4}.
For the fits of $\sigma_{8}$ and $\Omega_{\rm m}$ we also use
the mean of $P(v_{p})$ and the direct method; while for the velocity power spectrum fits
we use the median of $P(v_{p})$ (viz, $v_{i} = \text{Median}[P(v_{i})]$).
Here we have combined the constraints from different mock realisations.
Note for the separate fits using $\delta m$ and the PV we have used the same 
mock realisations.
 We interpret the
slight offset from the fiducial model (still within $1\sigma$) of the constraints found using $\delta m$ 
as simply a result of cosmic variance and covariance between mock realisations.

We conclude that for the constraints on $\sigma_{8}$ and $\Omega_{\rm m}$ using the mean, median and ML of $P(v_{p})$ and the direct method 
in the likelihood analysis all introduce a significant bias (i.e., $>2 \sigma$) in the final cosmological parameter values when 
considering a radial and angular halo distribution similar to 6dFGSv (and averaging over 8 realisations).
We find a similar, yet less significant, bias for the velocity power spectrum, given
the derived constraints are now consistent at the two sigma level.
As shown in the left panel of Fig. \ref{check_code:4}, the result is {\it more power}
relative to the fiducial cosmology
on the largest scales, which is consistent with a low bias 
in $\Omega_{\rm m}$. 
The non-Gaussian distributions 
imprint a bias in the mean radial velocity and therefore  
influence power on the largest scale.
Once a full sky survey is considered this effect is less severe as
the bias tends to averages out.

We test the sensitivity of the final constraints
to the process of marginalising over the zero-point. We find that the final results are 
reasonably insensitive to this procedure. As expected, the error in measurements on the 
largest scales is increased, which slightly weakens the constraints in the largest scale bin 
for the growth rate and velocity power spectrum measurements, and equivalently weakens the constraints 
on the matter density $\Omega_{\rm m}$. 

\section{Parameter fits to velocity data sets}
\label{sec:results}

In this section we present the results from the analysis of the 6dFGSv and low-$z$ SNe 
peculiar velocity surveys. Analysing the fluctuations in the measured PVs
and their correlations (as a function of their spatial separation) we 
are able to derive constraints on the following: the cosmological parameters 
$\Omega_{\rm m}$ and $\sigma_{8}$ (Section \ref{sec:con_omega});
the amplitude of the velocity power spectrum, ${\mathcal P}_{v v}(k) \equiv P_{\theta \theta}(k)/k^{2}$
in a series of (five) $\Delta k \sim 0.03 h{\rm Mpc^{-1}}$ bins (Section \ref{par:2}); the scale-dependent 
normalized growth rate of structure, $f\sigma_{8}(z=0,k)$,
in a series of (five) $\Delta k \sim 0.03h{\rm Mpc^{-1}}$ bins (Section \ref{sec:growth}); and the scale-independent
growth rate of structure, $f\sigma_{8}(z=0)$ (Section \ref{sec:growth}). 
All the constraints given are at a redshift $z\sim 0$. We emphasize that,
because we have not included any information from the local density field, 
as inferred by the local distribution of galaxies, the 
results presented here do {\it not} rely on any assumptions
about galaxy bias. Additionally, here we are working solely within the 
standard $\Lambda$CDM model.

For sections \ref{sec:con_omega}, \ref{par:2}, \ref{sec:growth} we 
give the results derived when analysing the individual surveys separately.
Comparing the results from different PV surveys allows one to check
for systematic effects.
When combining the PV surveys we consider two different approaches; both
introduce extra degrees of freedom that allow the relative `weight' of each 
sample to vary in the likelihood calculation. 
Firstly, we introduce a free parameter $\sigma_{{\rm v}}$
to each survey, this term accounts for non-linear velocity dispersion.
Secondly, we allow the relative weight of each survey to be varied by the 
use of a matrix hyper-parameter method (introduced in Section \ref{sec:combsurveys}). 
In this case we fix the $\sigma_{\rm v}$
values of both surveys to the maximum likelihood values found when analysing the 
surveys separately. The purpose of the hyper-parameter analysis is
to check the statistical robustness of our constraints. In the case
that the
hyper-parameter analysis is statistically consistent with the standard method 
of combining the surveys we quote the results from the standard method as
our final measurement.
The two PV samples we use for this analysis have significant overlap, therefore we 
expect the individual results to be highly correlated, given they
share the same cosmic variance. This limits the benefits from combining 
the samples. In addition complications arise when data points from each survey are
placed on the same grid point, as occurs when the velocity surveys are separately smoothed onto grids\footnote{
We treat these data points as if they were perfectly correlated in the full covariance matrix.}.

For all likelihood calculations in the following sections we marginalise 
over the unknown zero-point\footnote{We allow each survey to have different zero-point offsets for the marginalisation.}
(i.e., a monopole contribution to the velocity field).
The result of this process is that our constraints are not sensitive to the
uncertainties present in the determination of the zeropoint in PV surveys
and the assumptions required to determine the zeropoint.

\subsection{MCMC sampling strategy}
To sample the posterior distributions we use a python implementation of the
affine-invariant ensemble 
sampler for Markov Chain Monte Carlo (MCMC) {\tt MCMC-hammer} \citep{Foreman-Mackey:2012dq}. This technique was
introduced by \citet{Goodman:2010cr}.
We use the {\tt MCMC-hammer} algorithm because,
relative to the standard Metropolis--Hastings (M--H) algorithm 
the integrated autocorrelation time is lower and less `tuning' is required; specifically, only two parameters are required  
to tune the performance of the Markov chain, as opposed to $N [N +1]/2$ parameters in M--H, where $N$ is the 
dimension of the parameter space. 
Additionally the {\tt MCMC-hammer} 
algorithm is trivially parallelized using {\tt  MPI} and    
the {\it affine invariance} (invariance under linear transformations) 
property of this algorithm means
it is independent of covariances between parameters\footnote{No internal orthogonalisation of parameters is required.}
\citep{Foreman-Mackey:2012dq}.

We discard the first $20\%$ of each chain as `burn in' given that the sampling may be 
non-Markovian, while the convergence of each chain is assessed using the 
integrated autocorrelation time. From the samples we generate an
estimate of the posterior maximum-likelihood (ML) and median; given the 
posterior distributions of the parameters tend to be non-Gaussian, 
the $68\%$ confidence intervals we quote are found by calculating the
$34\%$ limits about the estimated median. In the case where we cannot 
quote a robust lower bound, when the probability distribution peaks near zero,
we quote $95\%$ upper limits. 

\subsection{Matter density and clustering amplitude\label{sec:con_omega}}

The base set of parameters we allow to vary in this analysis is $[\Omega_{\rm m}, \sigma_{8},\sigma_{\rm v}]$.
In the case where we combine PV surveys we consider two extensions to this base set. Firstly,
we include a free parameter modelling the non-linear velocity dispersion $\sigma_{\rm v}$ for 
each survey and therefore consider the set of parameters $[\Omega_{\rm m}, \sigma_{8},\sigma_{v,1},\sigma_{v,2}]$. 
Secondly, we fix the values for the velocity dispersion and introduce hyper-parameters, this gives the set
$[\Omega_{\rm m}, \sigma_{8},\alpha_{\rm 6dF},\alpha_{\rm SNe}]$.

For each likelihood evaluation of the cosmological parameters 
we must compute the corresponding velocity power spectrum. 
While the calculation of the velocity power spectrum in {\tt velMPTbreeze}
in significantly faster than previous RPT calculations, it remains too slow 
to embed directly in MCMC calculations. Therefore the approach we take 
here is to pre-compute a grid of velocity power spectra then use
a bilinear interpolation between the grid points to estimate the power 
spectra. 

Using {\tt velMPTbreeze} we evaluate a grid of velocity power 
spectra; we use the range $\Omega_{\rm m} = [0.050,0.500]$ and $\sigma_{8} = [0.432,1.20]$,
which act as our priors.
We use step sizes of $\Delta \Omega_{\rm m} = 0.01$
and $\Delta \sigma_{8} = 0.032$. 
We do not investigate the region of parameter space where $\Omega_{\rm m} < 0.05$
as here the theoretical modelling of the velocity power spectrum becomes uncertain
as it becomes highly non-linear on very large scales.
The prior placed on all $\sigma_{\rm v}$ parameters is $\sigma_{\rm v}$=$[0, 1000] \rm{km~{s}^{-1}}$
and $\alpha_{i} = [0,10]$.
For each value of $\Omega_{\rm m}$ the matter transfer function needs 
to be supplied, to do this we use the {\tt CAMB}
software package \citet{Lewis:2000nx}.
The numerical integration over the velocity power spectrum
requires us to specify a $k$-range. Here we integrate 
over the range $k = [0.0005,0.15]h {\rm Mpc^{-1}}$.
We note that integrating to larger scales (i.e. smaller 
values of $k$) when computing the full covariance matrix 
has a negligible effect on the derived constraints. 
Additionally, for the constraints given in this section we smooth
the local velocity field with a gridding scale of $20h^{-1}\rm{Mpc}$.

\begin{figure*}
\centering
\includegraphics[width=10cm]{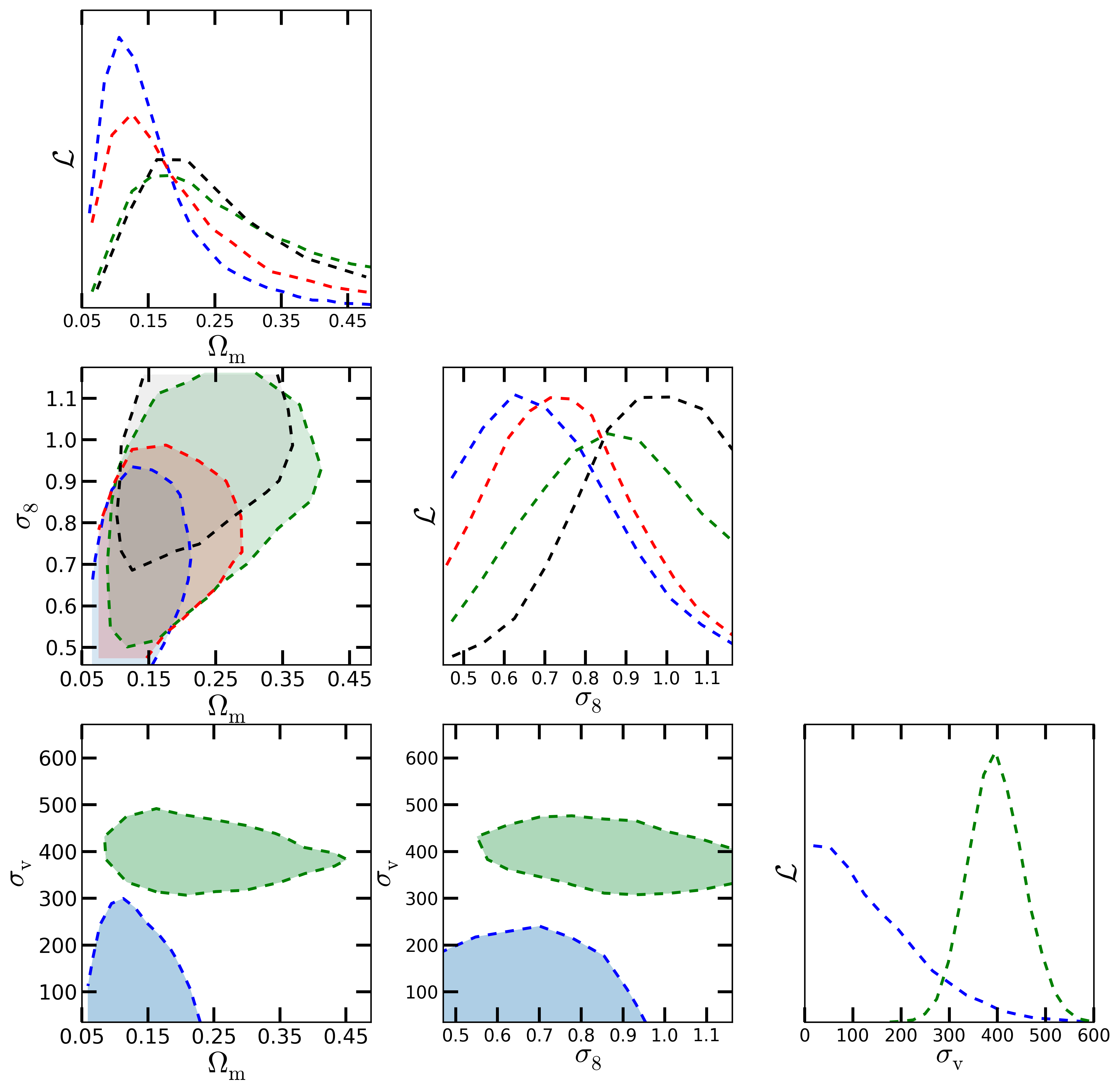}
\caption{68 \% confidence intervals for the matter density $\Omega_{\text{m}}$, $\sigma_{8}$ and the non-linear velocity dispersion $\sigma_{\rm v}$. 
Results are shown for 6dFGSv (blue), the SN sample (green), the combined analysis (red) and the combined hyper-parameter analysis (black).
The $\sigma_{\rm v}$ constraints from the combined analysis are very similar to the individual constraints hence we do not add them here.}
\label{plot:mcmc_omega}
\end{figure*}

The constraints for the parameters are shown in Fig. \ref{plot:mcmc_omega} 
and the best-fit values and 68\% confidence regions are given in Table \ref{tab2}.
Using only the 6dFGSv sample we determine $\Omega_{\text{m}} = 0.136^{{+0.07}}_{{-0.04}}$
and $\sigma_{8} = 0.69^{{+0.18}}_{{-0.14}}$, and for the SNe velocity sample
we determine $\Omega_{\text{m}} = 0.233^{{+0.134}}_{{-0.09}}$
and $\sigma_{8} = 0.86 \pm 0.18$. The results show that the two PV samples 
are consistent with each other and given the size of the errors we do not find
a strong statistical tension (less than $2\sigma$) with the parameter values reported by {\it Planck}. 
Combining the two PV surveys we determine $\Omega_{\text{m}} = 0.166^{ {+0.11}}_{ {-0.06}}$
and $\sigma_{8} = 0.74 \pm 0.16$; similarly 
we find no strong statistical tension with {\it Planck}. 
For the matrix hyper-parameter analysis we 
find $\alpha_{\rm 6dF} = 1.23 \pm 0.05$, $\alpha_{\rm SNe} = 0.87\pm 0.08$,
$\Omega_{\text{m}} = 0.228^{ {+0.12}}_{ {-0.08}}$ and $\sigma_{8} = 0.96 ^{ {+0.14}}_{ {-0.16}}$;
although the constraints from the hyper-parameters are best fit with the slightly 
higher $\sigma_{8}$ value, we find the results from the hyper-parameter analysis are statistically  
consistent with the previous constraints, as shown in 
Fig. \ref{plot:mcmc_omega}.

The constraints on $\Omega_{\rm m}$ and $\sigma_{8}$ outlined in this section,
while not competitive in terms of statistical uncertainty to other cosmological probes,
do offer some insight. In contrast to most methods to determine the matter density,
$\Omega_{\rm m}$, constraints from PV do not result from determining properties of the global statistically 
homogeneous universe (geometric probes); the constraints arise from the dependence of
the clustering properties of dark matter on $\Omega_{\text{m}}$. The 
consistency between these probes is an strong test of the cosmological model.

\begin{table*}
\begin{center}
\caption{Derived cosmological parameter values for $\Omega_{\rm m}$ and $\sigma_{8}$ plus the derived value for the non-linear 
velocity dispersion $\sigma_{\rm v}$ and the hyper-parameters $\alpha_{\rm 6dF}$ and $\alpha_{\rm SNe}$. 
Parameters not allowed to vary are fixed at their {\it Planck} ML values. 
Columns 2 and 3 give results from the 6dFGSv survey data alone. Columns 4 and 5 give results from the SNe sample data alone. 
For columns 6 and 7 we give the results combining both surveys; and for columns 8 and 9 we 
give the results combining both surveys using a matrix hyper-parameter analysis.
Note the hyper-parameters are only given for columns 8 and 9 as they are not included 
in the other analysis. All varied parameters are given flat priors.}

\label{tab2}
\begin{tabular}{@{}cccrccrccrccrc@{}}\toprule
\\ [-1.5ex]
&& 6dFGSv && & SNe &&& 6dFGSv + SNe (Norm) && & 6dFGSv + SNe (Hyp) \\
\cmidrule{2-3} \cmidrule{5-6}  \cmidrule{8-9} \cmidrule{11-12} 
Parameter & ML & \ && ML & Median  && ML & Median && ML & Median \\
& & [$68$ \% limits ] &&& [$68$ \% limits] &&& [$68$ \% limits ] &&& [$68$ \% limits] \\
\hline\\[-2ex]
$\Omega_{\rm m}$
& 0.103
& $0.136^{ {+0.07}}_{ {-0.04} }$
&
& 0.169
&$0.233^{ {+ 0.134}}_{ {-0.09}}$
&
& 0.107 
& $0.166^{ {+0.11}}_{ {-0.06}}$
&
& 0.183
& $0.228^{ {+0.12}}_{ {-0.08}}$
& \\[+1ex]
$\sigma_{8}$
& 0.66
& $0.69^{ {+0.18}}_{ {-0.14}}$
&
& 0.89
&$0.86 \pm 0.18$
&
& 0.73 
& $0.74 \pm 0.16 $
&
& 1.06
& $0.96 ^{ {+0.14}}_{ {-0.16}}$
& \\[+1ex]
$\sigma_{\rm v}$[km/s]
& 32.7
&114$^{ {+245}}_{ { }}$
&
& 388
& $395 ^{ {+54}}_{ {-58}}$
&
&--
&--
&
& --
& --
& \\[+1ex]
$\alpha_{\rm 6dF}$
& --
& --
& 
&--
&--
&
&--
&--
&
& 1.22
&$1.23 \pm 0.05$
& \\[+1ex]
$\alpha_{\rm SNe}$
& --
& --
&
& --
& --
&
&--
&--
&
&0.86
&$0.87\pm 0.08$
& \\
\bottomrule
\end{tabular}
\end{center}
\end{table*}
\begin{figure*}
\centering
 \includegraphics[width=15cm]{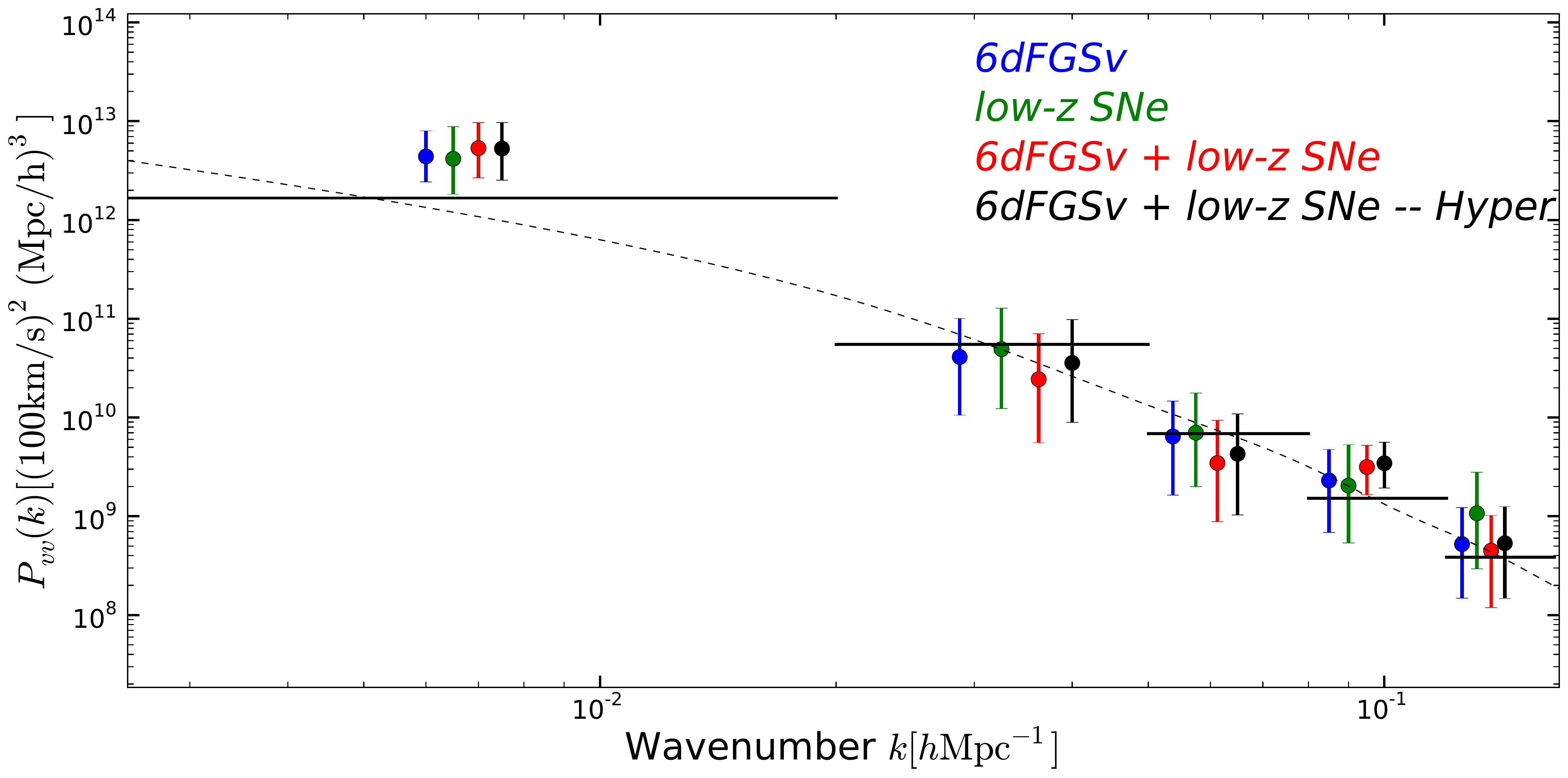}
\caption{ 68\% confidence intervals for the amplitude parameters $A_{i}$ scaled by the mean power within each bin
for the 6dFGSv data, SNe data and the combined constraint. The thick blue lines give the mean power in each bin in the 
fiducial cosmology calculated using Eq.~(\ref{v_mean}).
The black dashed line shows the velocity power spectrum ${\mathcal P}_{ v v }(k)$ calculated assuming the {\it Planck} cosmology. 
The circle symbols here give the median of the posterior distribution.}
\label{plot:power}
\end{figure*}
\begin{table*}
\begin{center}
\caption{Constraints on the velocity power spectrum amplitude parameters $A_{i}$
plus the value of the non-linear 
velocity dispersion $\sigma_{\rm v}$ and the hyper-parameters $\alpha_{\rm 6dF}$ and $\alpha_{\rm SNe}$. 
Parameters not allowed to vary are fixed at their {\it Planck} ML values. 
Columns 2 and 3 give results from the 6dFGSv survey data alone. Columns 4 and 5 give results from the SNe sample data alone. 
For columns 6 and 7 we give the results combining both surveys; and for columns 8 and 9 we 
give the results combining both surveys using a matrix hyper-parameter analysis. 
All varied parameters are given flat priors.}
\label{tabA}
\begin{tabular}{@{}cccrccrccrccr@{}}\toprule
\\ [-1.5ex]
&& 6dFGSv   & && SNe &&& 6dFGSv + SNe (Norm)   & &&  6dFGS + SNe (Hyp)  \\
\cmidrule{2-3} \cmidrule{5-6} \cmidrule{8-9} \cmidrule{11-12} 
Parameter & ML & Median && ML & Median  && ML & Median && ML & Median & \\
& & [$68$ \% limits ] && &[$68$ \% limits ] &&& [$68$ \% limits ] && &[$68$ \% limits ] & \\
\hline\\[-2ex]
$A_{1}(k_{1})$
&1.98
&$2.64^{ {+ 2.15}}_{ {-1.18}}$
&
&1.62
&$2.50^{ {+2.80}}_{ {-1.40}}$
&
&2.43
&$3.20^{ {+2.62}}_{ {-1.60}}$
&
&2.22
&$3.17^{ {+2.64}}_{ {-1.65}}$
& \\[+1ex]
$A_{2}(k_{2})$
&0.20
&$0.74^{ {+1.08}}_{ {- 0.55}}$
&
&0.25
&$0.89^{ {+1.43}}_{ {-0.67}}$
&
&0.14
& $0.44^{ {+0.84}}_{ {-0.34}}$
&
&0.26
&$0.65^{ {+1.13}}_{ {-0.49}}$ 
& \\[+1ex]
$A_{3}(k_{3})$
&0.20
&$0.94^{ {+1.20}}_{ {-0.70 }}$
&
&0.57
&$1.0^{ {+1.55}}_{ {-0.73}}$
&
&0.13
&$0.50^{ {+0.86}}_{ {-0.38}}$
&
&0.27
&$0.63^{ {+0.96}}_{ {-0.48}}$ 
& \\[+1ex]
$A_{4}(k_{4})$
&0.27
&$1.51^{ {+ 1.61}}_{ {-1.06}}$
&
&0.43
& $1.34^{ {+2.14}}_{ {-0.99}}$
&
&1.52
&$2.07^{ {+1.37}}_{ {-0.98}}$
&
&1.89
&2.26$^{ {+1.43}}_{ {-0.99}}$   
& \\[+1ex]
$A_{5}(k_{5})$							
&0.30
& $1.36^{ {+1.84}}_{ {-0.98 }}$
&
&0.84
&$2.79^{ {+4.49}}_{ {-2.03}}$
&
& 0.38
&$1.17^{ {+1.48}}_{ {-0.86}}$				
&
&0.40
&$1.39^{ {+1.86}}_{ {-1.00}}$   		
& \\[+1ex]
$\sigma_{\rm v}$[km/s]
& 98.4
& $137.5 ^{ {+110}}_{ {-91 }}$
& 
& 372.8
& $ 365.2^{ {+43}}_{ {- 45}}$
&
& --
& --
&
& --
& --
& \\[+1ex]
$\alpha_{\rm 6dF}$
& --
& 
& 
& --
& --
&
& --
& --
&
&1.198
&$1.189 \pm 0.034$
& \\[+1ex]
$\alpha_{\rm SNe}$
& --
& --
& 
& --
& --
&
& --
& --
&
& 0.940
&$ 0.980^{ {+0.104}}_{ {-0.091}}$
&
\\[+1.0ex]
\bottomrule
\end{tabular}
\end{center}
\end{table*}
\subsection{Velocity power spectrum\label{par:2}}

Analysing the surveys individually we consider the base parameter 
set $[A_{1}(k_{1}),A_{2}(k_{2}),A_{3}(k_{3}),A_{4}(k_{4}),A_{5}(k_{5}),\sigma_{\rm v}]$. Each $A_{i}$ parameter (defined in Eq.~(\ref{eqn:defA}))
acts to scale the amplitude of the velocity power spectrum, ${\mathcal P}_{v v}(k)$, over a 
specified wavenumber range given by $k_{1} \equiv [0.005,0.02]$,
$k_{2} \equiv [0.02,0.05]$, $k_{3} \equiv [0.05,0.08]$, $k_{4} \equiv [0.08,0.12]$ and
$k_{5} \equiv [0.12,0.150]$.
When combining samples we consider the parameter sets 
$[A_{1},A_{2},A_{3},A_{4},A_{5},\sigma_{v,1},\sigma_{v,2}]$ and 
$[A_{1},A_{2},A_{3},A_{4},A_{5},\alpha_{\rm 6dF},\alpha_{\rm SNe}]$.
We use a flat prior on the amplitude parameters, $A_{i} = [0,100]$, 
and the hyper-parameters $\alpha_{i} = [0,10]$.

The constraints for the amplitude of the velocity power spectrum are shown in Fig. \ref{plot:power} 
and the best-fit values and $68\%$ confidence regions are given in Table \ref{tabA}. 
The deviation between the ML values 
and median values (as shown in Table \ref{tabA}) is caused by the skewness of the distributions and the physical requirement 
that $A_{i} >  0$. This requirement results in a cut-off to the probability distribution that becomes more significant as
the size of the errors increases.
Therefore we caution that defining 
a single best-fitting value from the distribution requires subjective choices; 
note this is {\it not} the case for the growth rate constraints as shown 
in the next section. The fiducial power in each Fourier bin is 
consistent with that expected in our fiducial cosmological 
model assuming the best-fitting Planck parameters.

\subsection{Scale-dependent growth rate\label{sec:growth}}

We consider the results outlined in this section the most significant component of this work. 
We present the first measurement of a scale-dependent growth rate which includes
the largest-scale growth rate measurement to date (viz., length scales greater than $300h^{-1}\rm{Mpc}$).
Additionally, we present a redshift zero measurement of the growth rate that is {\it independent}
of galaxy bias and accurate to $\sim 15 \%$. Comparing this result to that obtained from 
the RSD measurement of 6dFGS \cite[i.e.,][]{Beutler:2012fk} allows one to 
test the systematic influence of galaxy bias, a significant source 
of potential systematic error in RSD analysis.

Analysing the surveys individually we consider two parameter sets: firstly we determine the 
growth rate in the scale-dependent bins defined above constraining the parameter set 
$[ f\sigma_{8} (k_{i}), \sigma_{\rm v}]$ ($i =1..5$);
secondly we fit for a single growth rate measurement
$[ f\sigma_{8} (z=0), \sigma_{\rm v}]$.
When combining data sets we consider the extensions to the base 
parameter set
$+[\sigma_{\rm v,1},\sigma_{\rm v,2}]$, and $+[ \alpha_{\rm 6dF},\alpha_{\rm SNe}]$ and 
use a smoothing length of $10h^{-1}{\rm Mpc}$. 
We fix the shape of the fiducial velocity power spectrum $\Omega_{\rm m}$
to the {\it Planck} value. By separating the power spectrum into wavenumber bins we expect that our final 
constraints are relatively insensitive to our choice of $\Omega_{\rm m}$. Varying $\Omega_{\rm m}$
generates a $k$-dependent variation in the power spectrum over very large scales; considering 
small intervals of the power spectrum this $k$-dependence is insignificant and to 
first order the correction to a variation in $\Omega_{\rm m}$ is simply a change 
in amplitude of the power spectrum,
which we allow to vary in our analysis.

\begin{figure*}
\centering
\includegraphics[width=15cm]{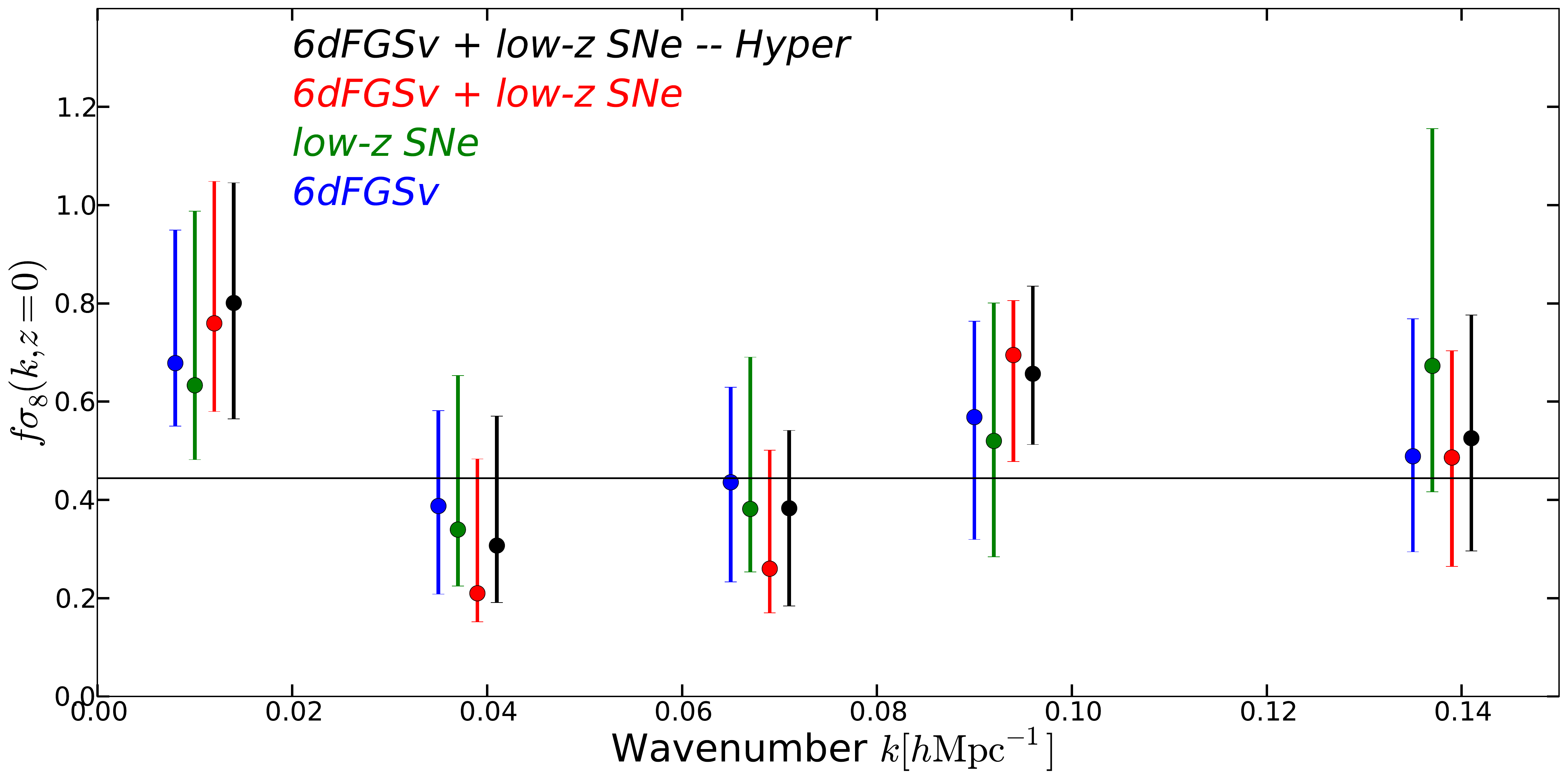}
\caption{68\% confidence intervals for the normalized {\it scale-dependent} growth rate $f(z=0,k)\sigma(z=0)$
in 5 different bins in Fourier space.
The thick black line is the prediction found assuming the fiducial {\it Planck} cosmology. For each $k$-bin we plot 
the results from 6dFGSv, the SNe sample and the combined constraint.
The bin intervals used here are given in Section \ref{par:2}. The largest scale bin 
corresponds to length scales  $> 300h^{-1}\rm{Mpc}$.
The circle symbols give the ML of the posterior distribution.}
\label{plot:growth5}
\end{figure*}

We first consider the scale-dependent constraints which are shown in Fig. \ref{plot:growth5};
with the best-fit and 68\% confidence internals given in Table \ref{tab1} and the full probability distributions in Fig.~\ref{plot:growth5_dis}.
For 6dFGSv we determine:
$f\sigma_{8}(k_{i})=[
0.72^{ {+0.17}}_{ {-0.23}},
0.38^{ {+0.17}}_{ {-0.20}}, 
0.43^{ {+0.20}}_{ {-0.20}}, 
0.55^{ {+0.22}}_{ {-0.23}}, 
0.52^{ {+0.25 }}_{ {-0.22}}]$. 
For the SNe velocity sample we have:
$f\sigma_{8}(k_{i})=[
0.70^{ {+0.29}}_{ {-0.22}},
0.42^{ {+0.23}}_{ {-0.19}},
0.45^{ {+0.24}}_{ {-0.20}},
0.51^{ {+0.29}}_{ {-0.23}}, 
0.74^{ {+0.41 }}_{ {-0.33}}]$.
As shown in Table \ref{tab1} the constraints on $\sigma_{\rm v}$ from 6dFGSv are very weak relative to the 
constraints from the SNe sample. The reason the $\sigma_{\rm v}$ parameter is much lower (and has a larger uncertainty) for the 6dFGSv 
sample relative to the SNe sample
is that the gridding has a stronger effect
for the 6dFGSv sample given the higher number density. 
This significantly reduces the contribution of non-linear velocity dispersion to the likelihood
and hence increases the final uncertainty. 
{In addition, we note that the magnitude of $\sigma_{\rm v}$ will be dependent on the 
mass of the dark matter halo that the galaxy resides in. The halo mass may vary between PV surveys,
therefore, causing $\sigma_{\rm v}$ to vary between PV surveys.}

The results (again) show that the
two survey are consistent with each other, viz., they 
are within one standard deviation of 
each other for all growth rate measurements. 
We detect no significant fluctuations from 
a scale-independent growth rate as predicted by the standard $\Lambda$CDM
cosmological model.
Although the power in the largest-scale Fourier bin is high,
it is consistent with statistical fluctuations.
When combining both the 6dFGSv sample and the SNe velocity sample we find (no hyper-parameters):
$f\sigma_{8}(k_{i})=[
0.79^{ {+0.21}}_{ {-0.25}},
0.30^{ {+0.14}}_{ {-0.19}}, 
0.32^{ {+0.19}}_{ {-0.15}}, 
0.64^{ {+0.17}}_{ {-0.16}}, 
0.48^{ {+0.22 }}_{ {-0.21}}]$.
We find no significant departure from the predictions
of the standard model. 

\begin{figure*}
\centering
\includegraphics[width=15cm]{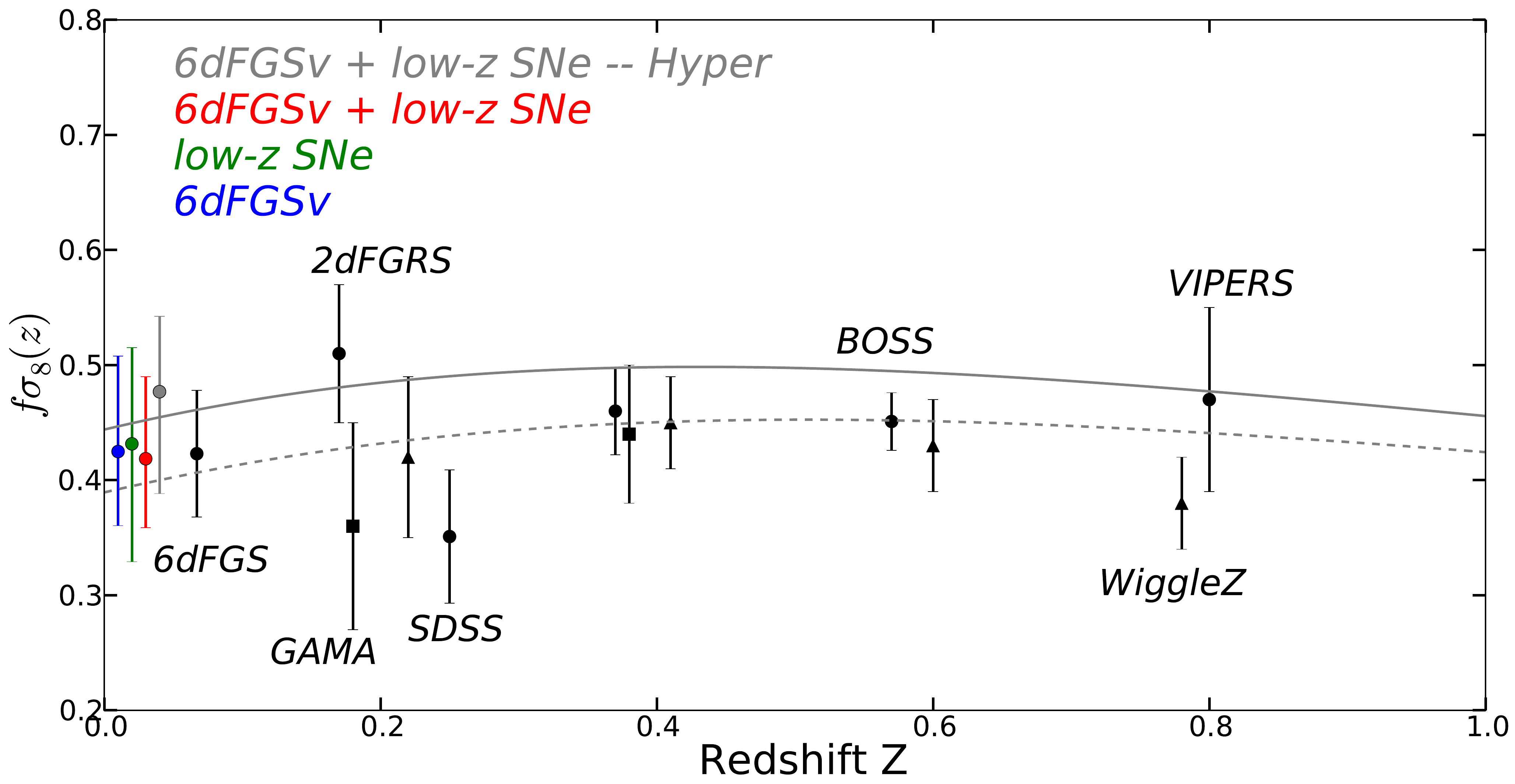}
\caption{68\% confidence intervals for the normalized growth rate $f(z=0)\sigma(z=0)$
averaging over all scales.The solid black line gives the theoretical prediction 
for $f\sigma_{8}(z)$ assuming the {\it Planck} cosmology and the 
dashed-black line gives the prediction assuming the {\it WMAP} cosmology. The 
redshift separation of the PV measurements (coloured points) is simply to avoid overlapping data
points; the redshift of the green data point gives the redshift of all the points.
We compare our PV measurements to previous constraints from redshift-space distortion 
measurements from the 6dFGS, 2dFGRS, GAMA, WiggleZ, SDSS LRG, BOSS CMASS and VIPERS surveys
given by the black points
\protect\citep{Beutler:2012fk, Hawkins:2003fk, 2011MNRAS.415.2876B, Blake:2013hc, Samushia:2013ys, Torre:2013fu}.}
\label{plot:growth1}
\end{figure*}

\begin{figure}
\centering
\includegraphics[width=9cm]{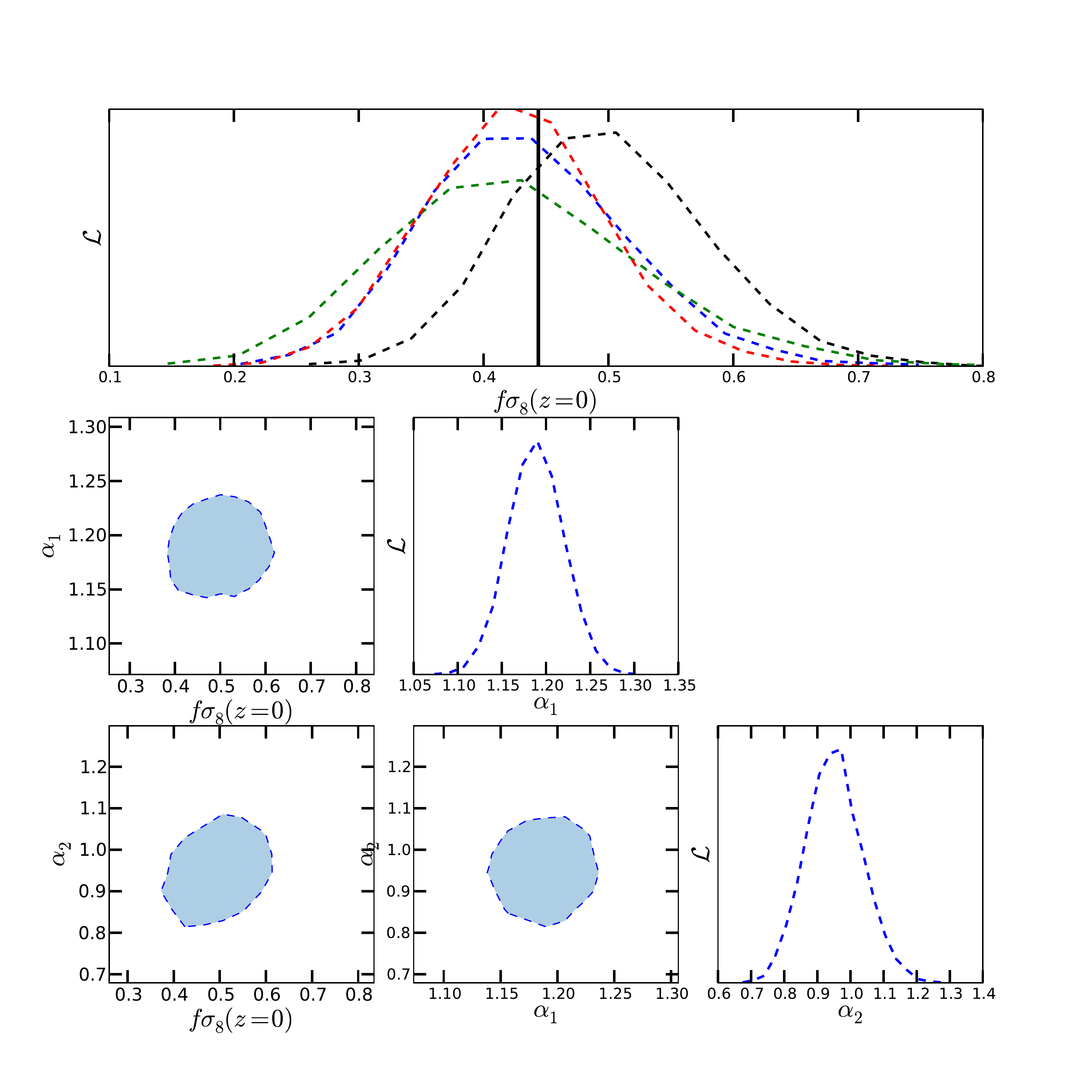}
\caption{Posterior distributions for the (scale averaged) growth rate of structure $f\sigma_{8}(z=0)$ for 6dFGSv (blue), SNe (green),
combining samples (red) and for the hyper-parameter analysis (black). The posterior distributions are also given for the 
hyper-parameters $\alpha_{\rm 6dF}$ and $\alpha_{\rm SNe}$. The prediction for the growth rate of structure assuming a fiducial {\it Planck}
cosmology is given by the solid black line.}
\label{plot:growth5_prob}
\end{figure}

\begin{figure*}
\centering
\includegraphics[width=12cm]{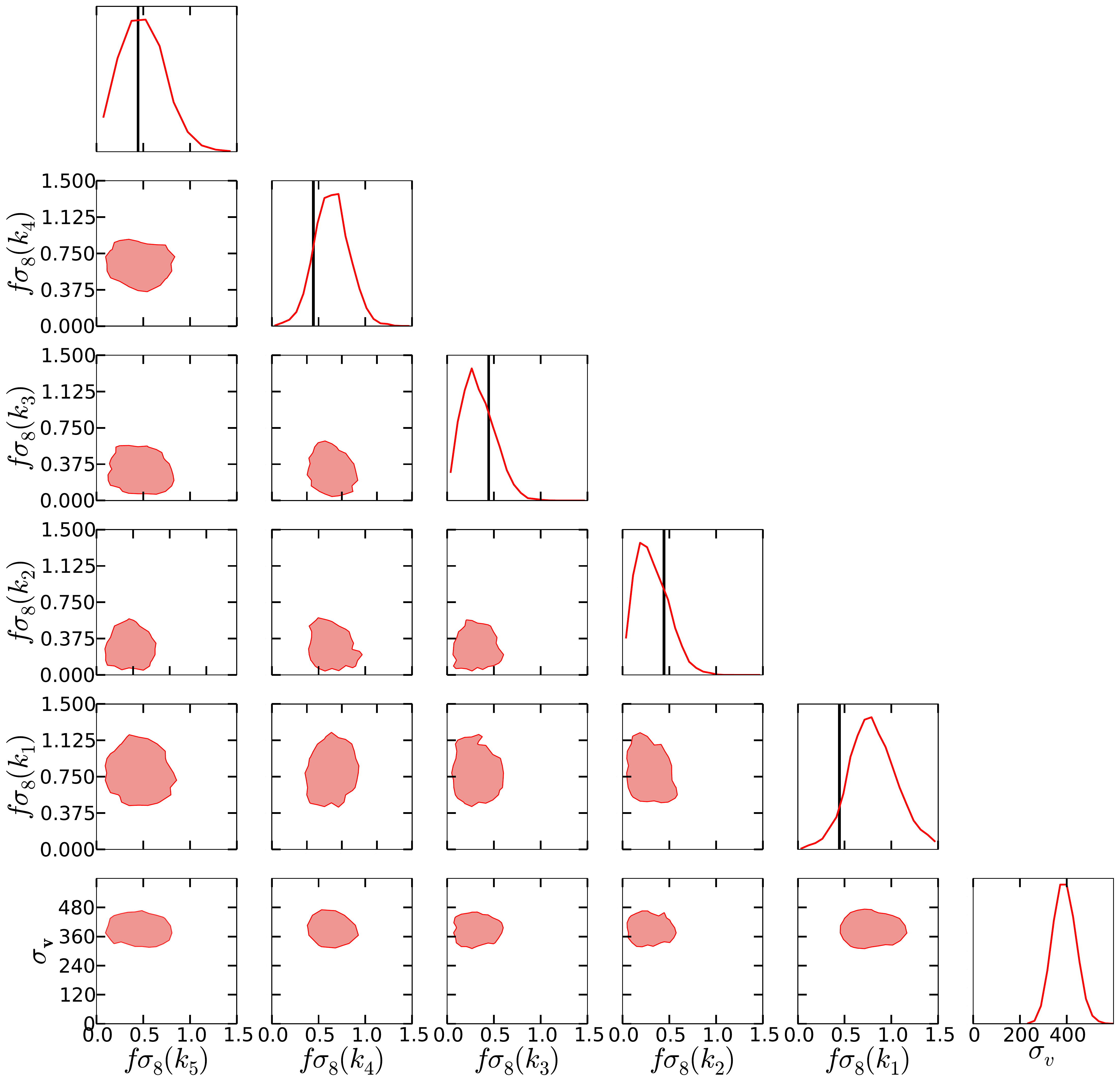}
\caption{68\% confidence intervals for the normalized growth rate $f(k,z=0)\sigma(z=0)$ for the combined constraints (using no hyper-parameters).
The prediction for the growth rate of structure assuming a fiducial {\it Planck}
cosmology is given by the solid black line.}
\label{plot:growth5_dis}
\end{figure*}

We next fit for a scale-independent growth rate by scaling the fiducial 
power spectrum across the full wavenumber range.
The measurements of a scale-independent growth rate of structure
are given in Fig. \ref{plot:growth1}. Here we also compare with
previously published results from RSD measurements and the 
predictions from the assumed fiducial cosmology.
The best-fit values and 68\% confidence intervals are given at the bottom 
of Table \ref{tab1}. We also plot the full probability distributions in Fig.~\ref{plot:growth5_prob},
in addition to the results from the hyper-parameter analysis.
For 6dFGSv, the SNe velocity sample and 6dFGSv+ SNe (with no hyper-parameters)
we determine, respectively, $f\sigma_{8}(z) = 
[ 0.428^{ {+0.079}}_{ {-0.068}}, 0.417^{ {+0.097}}_{ {-0.084}}, 0.418\pm 0.065]$. 
The measurements of the growth rate all show consistency with the 
predictions from the fiducial model as determined by {\it Planck}. Specifically, the best fitting {\it Planck
} parameters predict $f\sigma_{8}(z=0) = 0.443$. 
In addition we find consistency with the measurement of
the growth rate of structure from the RSD analysis of the 6dFGS 
(see Fig. \ref{plot:growth1}) \citep[]{Beutler:2012fk}.	

For the hyper-parameter analysis the results for the scale-dependent and scale-independent measurements
are indistinguishable. We determine $\alpha_{\rm 6dF} =1.189 \pm 0.034$ and $\alpha_{\rm SNe} = 0.980^{ {+0.104}}_{ {-0.091}}$;
the results for both analysis have been included in Fig. \ref{plot:growth5} and Fig. \ref{plot:growth1}.
We find that, while there is a slight shift in the best-fit values, the hyper-parameter analysis gives 
results statistically consistent with the previous results; for the scale-independent measurements this 
is best shown in Fig. \ref{plot:growth5_prob}.

\begin{table*}
\begin{center}
\caption{Constraints on the growth rate as a function of scale
and independent of scale (final row)
plus the value of the non-linear 
velocity dispersion $\sigma_{\rm v}$ and the hyper-parameters $\alpha_{\rm 6dF}$ and $\alpha_{\rm SNe}$. 
Columns 2 and 3 give results from the 6dFGSv survey data alone. Columns 4 and 5 give results from the SNe sample data alone. 
For columns 6 and 7 we give the results combining both surveys; and for columns 8 and 9 we 
give the results combining both surveys using a matrix hyper-parameter analysis.}
\label{tab1}
\begin{tabular}{@{}cccrccrccrccr@{}}\toprule
\\ [-1.5ex]
&& 6dFGSv   & && SNe &&& 6dFGSv + SNe (Norm)   & &&  6dFGS + SNe (Hyp)  \\
\cmidrule{2-3} \cmidrule{5-6} \cmidrule{8-9} \cmidrule{11-12} 
Parameter & ML & Median && ML & Median  && ML & Median && ML & Median & \\
& & [$68$ \% limits ] && &[$68$ \% limits ] &&& [$68$ \% limits ] && &[$68$ \% limits ] & \\
\hline \\[-2ex]
$f \sigma_{8}(k_{1})$
& 0.68
& $0.72^{ {+0.17}}_{ {-0.23}}$
&
&0.63
&$0.70^{ {+0.29}}_{ {-0.22}}$
&
& 0.76
& $0.79^{ {+0.21}}_{ {-0.25}}$
&
&0.79
&$0.80^{ {+0.23}}_{ {-0.25}}$ 
& \\ [+1ex]
$f \sigma_{8}(k_{2})$
& 0.39
& $0.38^{ {+0.17}}_{ {-0.20}}$
&
&0.34
& $0.42^{ {+0.23}}_{ {-0.19}}$ 
&
& 0.21
& $0.30^{ {+0.14}}_{ {-0.19}}$
&
& 0.31
&$0.36^{ {+0.17}}_{ {-0.21}}$ 
& \\ [+1ex]
$f \sigma_{8}(k_{3})$
&0.44
&$0.43^{ {+0.20}}_{ {-0.20}}$
&
& 0.38
&$0.45^{ {+0.24}}_{ {-0.20}}$ 
&
&0.260
&$0.32^{ {+0.19}}_{ {-0.15}}$
&
& 0.38
&$ 0.35^{ {+0.17}}_{ {-0.19}}$ 
& \\ [+1ex]
$f \sigma_{8}(k_{4})$
& 0.57
& $0.55^{ {+0.22}}_{ {-0.23}}$
&
&0.52
&$0.51^{ {+0.29}}_{ {-0.23}}$ 
&
& 0.69
& $0.64^{ {+0.17}}_{ {-0.16}}$
&
& 0.66
& $0.66^{ {+0.17}}_{ {-0.19}}$  
& \\ [+1ex]
$f \sigma_{8}(k_{5})$
&0.49
& $0.52^{ {+ 0.25}}_{ {-0.22}}$
&
& 0.67
& $0.74^{ {+0.41}}_{ {-0.33}}$ 
&
& 0.49
& $0.48^{ {+ 0.22}}_{ {-0.21}}$ 
&
& 0.53
&$0.52^{ {+0.15}}_{ {-0.17}}$ 
& \\ [+1ex]
$\sigma_{\rm v}$ [km/s]
& 98.4
& $137.5 ^{ {+110}}_{ {-91 }}$
& 
& 372.8
& $ 365.2^{ {+43}}_{ {- 45}}$
&
& --
& --
&
& 98.4
& 372.8
& \\[+1ex]
$\alpha_{\rm 6dF}$
& --
& --
& 
& --
& --
&
& --
& --
&
&1.198
&$1.189 \pm 0.034$
& \\[+1ex]
$\alpha_{\rm SNe}$
& --
& --
& 
& --
& --
&
& --
& --
&
&0.940
&$ 0.980^{ {+0.104}}_{ {-0.091}}$
&\\ [+1.0ex]
\hline \\ [-1.5ex]
$ f \sigma_{8}(z =0)$
& 0.424
& $0.428^{ {+0.079}}_{ {-0.068 }}$
&
& 0.432
& $0.417^{ {+0.097}}_{ {-0.084}}$
&
& 0.429
& $0.418\pm 0.065$
&
& 0.492
&$0.496^{ {+0.044}}_{ {-0.108}}$
& \\[+1.0ex]
\bottomrule
\end{tabular}
\end{center}
\end{table*}
\section{Discussion and conclusions}
\label{sec:con}

We have constructed 2-point statistics of the velocity field and tested the 
$\Lambda$CDM cosmology by using low-redshift 6dFGSv and 
Type-Ia supernovae data.
We summarise our results as follows:
\begin{itemize}
\item{We introduced and tested a new method to constrain the scale-dependence 
of the normalized growth rate using only peculiar velocity data.  Using this method we 
present the {\it largest-scale} constraint on the growth rate of structure to date. 
For length scales greater than $\sim 300h^{-1}\rm{Mpc}$ ($k < 0.02h{\rm Mpc^{-1}}$) we constrain the growth rate to $\sim 30\%$.
Specifically, we find for 6dFGSv, which provides our best constraints, 
$f\sigma_{8}(k<0.02h\rm{Mpc}^{-1}) = 0.72^{+0.17}_{- 0.23}$. This result is 
consistent with the standard model prediction of $f\sigma_{8}(z=0)= 0.4439$, albeit higher than expected.}
\vspace{0.10cm}
\item{Examining the scale-dependence of the growth rate of structure at $z=0$ we find the constraints 
$f\sigma_{8}(k_{i})=[
0.79^{ {+0.21}}_{ {-0.25}},
0.30^{ {+0.14}}_{ {-0.19}},
0.32^{ {+0.19}}_{ {-0.15}},
0.64^{ {+0.17}}_{ {-0.16}}, 
0.48^{ {+ 0.22}}_{ {-0.21}}]$
using the wavenumber ranges
$k_{1} \equiv [0.005,0.02]$, $k_{2} \equiv [0.02,0.05]$, $k_{3} \equiv [0.05,0.08]$, $k_{4} \equiv [0.08,0.12]$ and
$k_{5} \equiv [0.12,0.150]$. We find no evidence for a scale-dependence in the growth rate, 
which is consistent with the standard model. All the growth rate measurements are consistent with the fiducial {\it Planck} 
cosmology.}\vspace{0.10cm}
\item{Averaging over all scales we measure the growth rate to $\sim 15\%$
which is {\it independent} of galaxy bias. This 
result $f\sigma_{8}(z=0) = 0.418 \pm 0.065$ is consistent with the redshift-space distortion analysis
of 6dFGS which produced a measurement of $f\sigma_{8}(z) = 0.423 \pm 0.055$ \citep{Beutler:2012fk}, 
increasing our confidence in the modelling of galaxy bias. In addition this measurement is consistent 
with the constraint given by \citet{2012ApJ...751L..30H} of $f\sigma_{8} = 0.400 \pm 0.07$,
found by comparing the local velocity and density fields. In contrast to our
constraint this measurement is sensitive to galaxy bias and any systematic 
errors introduced during velocity field reconstruction.}\vspace{0.10cm}
\item{We also consider various other methods to constrain the standard model. We 
directly constrain the amplitude of the velocity power spectrum 
${\mathcal P}_{v v}(k)\equiv {\mathcal P}_{\theta \theta}(k)/k^2$ for the same scale 
range as specified above; we find that the predictions from two loop multi-point propagators assuming the 
{\it Planck} cosmology gives an accurate description of the measured velocity power spectrum. Specifically, the
derived amplitudes $A_{i}$ of the power spectrum of 4 bins are consistent with the fiducial cosmology at the $1\sigma$ level, 
and the largest scale bin
is consistent at the $2\sigma$ level. We can also compare these constraints to those given by 
\citet{Macaulay:2011bh}. Similarly to our results they 
found the amplitude of the matter power spectrum, determined using the composite sample 
of PVs, to be statistically consistent with the standard $\Lambda$CDM cosmology. In addition
they also find on the largest scales a slightly higher amplitude of the power spectrum
that expected in the standard model\footnote{Note
we cannot directly compare
these sets of results given different bin ranges were used.}.}\vspace{0.10cm}
\item{We show that when analysing PV surveys with velocities derived using the Fundamental Plane or the 
Tully-Fisher relation, one should perform the analysis using a variable that is a linear transformation of
$x = \log_{10}\left( D_{z} / D_{\rm H} \right)$. We show the intrinsic scatter is not Gaussian for the PV 
and this can significantly bias cosmological constraints. We show how the analysis can be reformulated 
using the variable $\delta m$, which removes the bias.}
\end{itemize}

With a large number of upcoming PV surveys, the prospect for
understanding how structure grows in the low-redshift universe is excellent.
Future work will move beyond consistency tests by adopting specific modified 
gravity models and phenomenological parametrisations, including measurements of
redshift-space distortions and by
self-consistently modifying the growth and 
evolutionary history of the universe.
This will allow a vast range of spatial and temporal scales 
to be probed simultaneously, providing a strong and unique test of 
the standard $\Lambda$CDM model, and 
perhaps even providing some insight on the so-far mysterious dark energy
component of the universe.

\section*{ACKNOWLEDGMENTS}

We are grateful for a very constructive referee report from Andrew Jaffe.
AJ, JK and CS are supported by the Australian Research Council Centre 
of Excellence for All-Sky Astrophysics (CAASTRO) through project number CE110001020.
CB acknowledges the support of the Australian Research Council through 
the award of a Future Fellowship. TMD acknowledges the support of the Australian 
Research Council through a Future Fellowship award, FT100100595.
This work was performed on the gSTAR 
national facility at Swinburne University of Technology. 
gSTAR is funded by Swinburne and the Australian 
Government’s Education Investment Fund.

\vspace{-0.3cm}

\bibliographystyle{mn2e}
\bibliography{6dfGSv_power}

\end{document}